%% file: main.tex
\documentclass[10pt,conference]{IEEEtran}
\usepackage{cite}
\usepackage{amsmath,amssymb,amsfonts}
\usepackage{algorithmic}
\usepackage{graphicx}
\usepackage{textcomp}
\usepackage{tikz}
\usepackage{xcolor}
\usepackage[hyphens]{url}
\usepackage{fancyhdr}
\usepackage[hidelinks]{hyperref}
\usepackage{hyperref}
\usepackage[caption=false,font=normalsize,labelfont=sf,textfont=sf]{subfig}
\usepackage{CJKutf8}
\usepackage{array}
\usepackage{stfloats}
\usepackage{url}
\usepackage{verbatim}
\usepackage{wasysym}
\usepackage{siunitx}
\usepackage{caption}
\usepackage{cuted}
\usepackage{balance}
\usepackage{makecell}
\usepackage{marvosym}

\graphicspath{{./figure/}}

\newcommand{\hpcayear}{2026}

\newsavebox{\circleone}
\savebox{\circleone}{%
  \tikz[baseline=-0.8ex]{\node[circle,fill=black,text=white,font=\normalsize,inner sep=0.8pt]{\textbf{1}};}%
}
\newcommand{\circledone}{\usebox{\circleone}}

\newsavebox{\circletwo}
\savebox{\circletwo}{%
  \tikz[baseline=-0.8ex]{\node[circle,fill=black,text=white,font=\normalsize,inner sep=0.8pt]{\textbf{2}};}%
}
\newcommand{\circledtwo}{\usebox{\circletwo}}

\newsavebox{\circlethree}
\savebox{\circlethree}{%
  \tikz[baseline=-0.8ex]{\node[circle,fill=black,text=white,font=\normalsize,inner sep=0.8pt]{\textbf{3}};}%
}
\newcommand{\circledthree}{\usebox{\circlethree}}

\newsavebox{\circlefour}
\savebox{\circlefour}{%
  \tikz[baseline=-0.8ex]{\node[circle,fill=black,text=white,font=\normalsize,inner sep=0.8pt]{\textbf{4}};}%
}
\newcommand{\circledfour}{\usebox{\circlefour}}

\newsavebox{\circlefive}
\savebox{\circlefive}{%
  \tikz[baseline=-0.8ex]{\node[circle,fill=black,text=white,font=\normalsize,inner sep=0.8pt]{\textbf{5}};}%
}
\newcommand{\circledfive}{\usebox{\circlefive}}

\newsavebox{\circlesix}
\savebox{\circlesix}{%
  \tikz[baseline=-0.8ex]{\node[circle,fill=black,text=white,font=\normalsize,inner sep=0.8pt]{\textbf{6}};}%
}
\newcommand{\circledsix}{\usebox{\circlesix}}

\newsavebox{\circleseven}
\savebox{\circleseven}{%
  \tikz[baseline=-0.8ex]{\node[circle,fill=black,text=white,font=\normalsize,inner sep=0.8pt]{\textbf{7}};}%
}
\newcommand{\circledseven}{\usebox{\circleseven}}

\begin{document}
%%%%%%%%%%%%%%%%%%%%%%%%%%%%%%%%%%%%%%%%
%%%%%%%%%%%%%% -- UPDATE -- %%%%%%%%%%%%%%%
\newcommand{\hpcasubmissionnumber}{399}

%\bstctlcite{IEEEexample:BSTcontrol}%参考文献设置

\title{Cohet: A CXL-Driven Coherent Heterogeneous Computing Framework with Hardware-Calibrated Full-System Simulation}
%%%%%%%%%%%%%%%%%%%%%%%%%%%%%%%%%%%%%%%%

%%%%%%%%%%%%%%%%%%%%%%%%%%%%%%%%%%%%%%%%
%%%%%%%% -- ONLY FOR CAMERA READY -- %%%%%%%%
\def\hpcacameraready{} % Uncomment to build camera-ready version
\newcommand{\hpcapubid}{0000--0000/00\$00.00}
\newcommand\hpcaauthors{Yanjing Wang\textsuperscript{$*$$\dagger$}\quad Lizhou Wu\textsuperscript{$*$$\dagger$}\quad Sunfeng Gao\textsuperscript{$\dagger$}\quad Yibo Tang\textsuperscript{$\dagger$}\quad Junhui Luo\textsuperscript{$\dagger$}\quad \\ Zicong Wang\textsuperscript{$\dagger$}\quad Yang Ou\textsuperscript{$\dagger$}\quad Dezun Dong\textsuperscript{$\dagger$}\quad Nong Xiao\textsuperscript{\Letter$\S$$\dagger$}\quad Mingche Lai\textsuperscript{\Letter$\dagger$}}

\newcommand\hpcaaffiliation{
\textsuperscript{$\dagger$}College of Computer Science and Technology, National University of Defense Technology, China \\
\textsuperscript{$\S$}School of Computer Science and Engineering, Sun Yat-sen University, China}

\newcommand\hpcaemail{\{wangyanjing, lizhou.wu, mingchelai\}@nudt.edu.cn, xiaon6@mail.sysu.edu.cn}

%%%%% -- ARTEFACT EVALUATION RESULTS -- %%%%%%
% Uncomment the following based on the badges that were awarded to this paper
%\def\aeopen{}           % The artifact is publically available
%\def\aereviewed{}     % The artefact has been reviewed
%\def\aereproduced{} % The results have been reproduced
%%%%%%%%%%%%%%%%%%%%%%%%%%%%%%%%%%%%%%%%

\input{hpca-template}

\maketitle

\makeatletter
\renewcommand{\@makefntext}[1]{\noindent #1} % 去掉脚注编号
\makeatother

\footnotetext{\textsuperscript{$*$}Co-first author. \quad \textsuperscript{\Letter}Co-corresponding author.}

%%%%%%%%%%%%%%%%%%%%%%%%%%%%%%%%%%%%%%%%
%%%%%%%% -- PAPER CONTENT STARTS -- %%%%%%%%%

\begin{abstract}
Conventional heterogeneous computing systems built on PCIe interconnects suffer from inefficient fine-grained host-device interactions and complex programming models. In recent years, many proprietary and open cache-coherent interconnect standards have emerged, among which compute express link (CXL) prevails in the open-standard domain after acquiring several competing solutions. Although CXL-based coherent heterogeneous computing holds the potential to fundamentally transform the collaborative computing mode of CPUs and XPUs, research in this direction remains hampered by the scarcity of available CXL-supported platforms, immature software/hardware ecosystems, and unclear application prospects. This paper presents Cohet, the first CXL-driven coherent heterogeneous computing framework. Cohet decouples the compute and memory resources to form unbiased CPU and XPU pools which share a single unified and coherent memory pool. It exposes a standard malloc/mmap interface to both CPU and XPU compute threads, which share a single per-process page table for user applications, leaving the OS dealing with smart memory allocation, page auto-migration, and management of heterogeneous resources. This design significantly simplifies heterogeneous parallel programming to a level comparable to homogeneous programming. To facilitate Cohet research, we also present a full-system cycle-level simulator named SimCXL, which is capable of modeling all CXL sub-protocols and device types. SimCXL has been rigorously calibrated against a real CXL testbed with various CXL memory and accelerators, showing an average simulation error of 3\%. Our evaluation reveals that CXL.cache reduces latency by 68\% and increases bandwidth by 14.4$\times$ compared to DMA transfers at cacheline granularity. Building upon these insights, we demonstrate the benefits of Cohet with two killer apps, which are remote atomic operation (RAO) and remote procedure call (RPC). Compared to PCIe-NIC design, CXL-NIC achieves a 5.5 to 40.2$\times$ speedup for RAO offloading and an average speedup of 1.86$\times$ for RPC (de)serialization offloading.
\end{abstract}

\vspace{-2pt}
\section{Introduction}
With the end of Dennard scaling and slowdown of Moore's law, homogeneous CPU architectures can no longer satisfy the explosive compute and memory demands of emerging data-driven workloads such as AI/ML and big data analytics~\cite{intel_accel_ecosystem}. This gap has spurred widespread adoption of heterogeneous systems that integrate CPUs with various XPUs (e.g., GPUs, SmartNICs/DPUs, and FPGAs) over high-speed interconnects such as PCIe to exploit their domain-specific accelerating capabilities \cite{tpuv4,dpu,ipu,apunet,dnn_processor,geeps_dl,azure_accelerated}, thus boosting overall computing performance. However, the conventional PCIe-based heterogeneous architecture suffers from two major limitations \cite{demystifying_type2,mozart,schuh_cc-nic_2024,intel_accel_ecosystem,introduction_cxl,cuda_um,gh100_evaluation}. 
First, PCIe favors occasional host-device transfers of bulky data sets, making it well suited to bandwidth-intensive coarse-grained workloads. In contrast, latency-sensitive applications that require frequent or fine-grained CPU-XPU interactions are severely hindered by PCIe’s high latency and poor bandwidth in small data transfers. 
Second, PCIe is a noncoherent interconnect that has to rely on complex software stacks and significant programmer efforts to explicitly or implicitly manage data movement between host and accelerator, leading to a cumbersome programming model and poor development productivity.

In recent years, both proprietary \cite{nvlink_c2c,infinity_fabric,intel_qpi_upi,intel_accel_ecosystem} and open \cite{CXL_spec,opencapi,genz,ccix} coherent interconnect technologies have emerged to address the limitations of PCIe. Proprietary solutions (e.g., NVIDIA's NVLink-C2C \cite{nvlink_c2c}, AMD's Infinity Fabric~\cite{infinity_fabric}, and Intel's UPI \cite{intel_qpi_upi} and CMI~\cite{intel_accel_ecosystem}) deliver impressive performance, but lack hardware flexibility, software ecosystem diversity, and user code portability.
Among open standards, CXL~\cite{CXL_spec} has become the de facto industry standard with over 240 active consortium members \cite{cxl_integrators} after consolidating competing proposals OpenCAPI~\cite{opencapi}, Gen-Z \cite{genz}, and CCIX~\cite{ccix}. Building upon the PCIe physical layer, CXL defines three sub-protocols: CXL.io, CXL.cache, and CXL.mem. To date, the majority of academic and industrial research works are focused on CXL-based disaggregated memory systems merely using CXL.mem \cite{demystifying_type3,exploring_cxl_asic,samsung_cxl_ssd,maruf2023tpp,pond,lia_cxl_llm}, which promise clear benefits in decoupling memory from compute and enabling independent scaling and sharing of memory resources. However, we contend that there is tremendous potential in exploiting CXL.cache together with CXL.mem to realize fully coherent heterogeneous computing. 
Such an approach can overcome the bottlenecks of the conventional heterogeneous architecture and fundamentally reshape the future cooperative computing paradigm.

Despite its promise, CXL-based coherent heterogeneous computing faces three key challenges. First, the CXL hardware and software ecosystem remains immature. On the hardware side, CXL memory expanders have only reached early commercial availability \cite{samsung_cmm_d,rambus,XConn,skhynix,Montage}, and CXL-supported accelerators remain largely confined to FPGA platforms~\cite{agilex,cxl_fpga_ip} in experimental testbeds. On the software side, the OS kernel lacks native support for CXL accelerators \cite{cxl_linux}, and higher-level heterogeneous programming frameworks must adapt to CXL cache coherence and memory semantics \cite{opencl,oneapi}. Second, academic researchers suffer from a dearth of realistic CXL platforms. Previous works often used a remote NUMA node as an emulator of CXL accelerators \cite{huang_hal_2024,tarot_row_hammer,schuh_cc-nic_2024}, which can introduce inaccuracies and misleading conclusions in some cases \cite{demystifying_type2}. Finally, owing to the lack of effective evaluation and comprehensive analysis of CXL accelerators, the specific application scenarios benefiting from coherent heterogeneous computing remain unexplored.

To address these challenges, this paper presents Cohet, the first CXL-driven \underline{co}herent \underline{het}erogeneous computing framework that is intended to deliver a software-hardware co-design solution targeting the open-source community. By interconnecting hosts and accelerators over CXL with cache-coherence, Cohet unifies their physical address spaces into a coherent memory pool. Cohet also provides a single shared memory view and a unified per-process page table to both CPUs and XPUs. The OS recognizes CPUs and XPUs as separate NUMA nodes, while providing standard malloc/mmap interfaces to programmers; memory allocation, page migration, and coherence management remain transparent. Moreover, we have developed SimCXL, an open-source full-system cycle-level simulator extended from gem5 that models all three CXL sub-protocols and three types of CXL devices. 
%We rigorously calibrated SimCXL against a real CXL-supported FPGA testbed, achieving simulation errors of less than 3\%. 
SimCXL has been rigorously calibrated against a real CXL-supported testbed with Intel FPGA accelerators and various memory expanders, achieving an average simulation error of 3\%. Our evaluation shows that at cacheline granularity, CXL.cache reduces latency by 68\% and increases bandwidth by 14.4$\times$ compared to DMA.
To demonstrate the benefits of Cohet, we present a CXL-NIC design with two acceleration case-studies on SimCXL: 
1) \textit{remote atomic operation} (RAO), a key building block for distributed parallel computing \cite{circustent,RAE,rao_sequencer,think_more_rdma,rdma_design,lock_2,ao_evaluating}, and 2) \textit{remote procedure call} (RPC), a fundamental inter-process communication layer in modern cloud microservices \cite{micro_server,rpcnic,dagger,google_2023,protoAcc,meta_2020}. Our experimental results suggest that CXL-based RAO outperforms its PCIe-based counterpart by 5.5 to 40.2$\times$ across different operation types, and that CXL-based RPC achieves an average 1.86$\times$ speedup in (de)serialization compared to PCIe-based RPC.

The main contributions are summarized as follows.
\begin{itemize}
\item A Cohet computing framework, which brings an optimized collaborative mode, decentralized control and synchronization, and a simplified programming model.
\item A full-system cycle-level simulator named SimCXL, with complete support for all three CXL sub-protocols and three types of CXL devices.
\item Comprehensive characterization of the latency and bandwidth of CXL.cache transactions; the measurements real new insights and are utilized to calibrate SimCXL.
\item CXL-NIC design with two killer-app demonstrations; RAO for distributed parallel computing and RPC for inter-process communication in cloud microservices, demonstrating substantial speedups over PCIe-NIC.
\end{itemize}
% \begin{itemize}
% \item We introduce Cohet, a CXL-driven coherent heterogeneous computing framework that overcomes the low collaboration efficiency of conventional heterogeneous architectures and simplifies the programming model.
% \item We develop SimCXL, a cycle-level, full-system simulator with complete support for all three CXL sub-protocols and three classes of CXL devices.
% \item We characterize the latency and bandwidth of CXL.cache on a real FPGA-based CXL testbed and rigorously calibrate SimCXL against it, achieving simulation errors of only 3\%.
% \item We reconstruct two key primitives in SimCXL under Cohet: RAO for distributed parallel computation and RPC for inter-process communication in cloud services, demonstrating substantial speedups over PCIe-based implementations.
% \end{itemize}
% =====================================================
\section{Background}
% =====================================================

\begin{figure}[tb]
\centering
\includegraphics[width=0.35\textwidth]{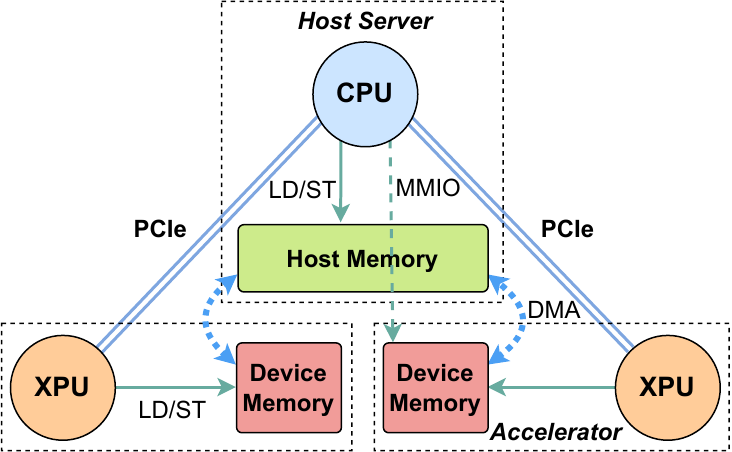}
\caption{PCIe-based  heterogeneous computing architecture with CPU-biased computing  and discrete physical memories.}
\label{Fig:PCIe_based}
\vspace{-15pt}
\end{figure}

\subsection{Conventional Heterogeneous Computing}
\label{Sec:Conventional_Heterogeneous_Computing}
%Today's heterogeneous computing systems integrate CPUs and XPUs such as GPUs, SmartNICs/DPUs, and FPGAs via high-speed interconnects (e.g., the industry standard PCIe) to exploit their domain-specific accelerating capabilities, boosting  the overall system performance. As the traffic backbone, the interconnect plays a key role in determining data movement and heterogeneous computing efficiency. 
Fig.~\ref{Fig:PCIe_based} shows a typical PCIe-based heterogeneous architecture, with its CPU-XPU interactive interface and collaborative computing mode elaborated as follows.

\subsubsection{CPU-XPU interactive interface} 
In PCIe-based systems, host and device memories reside in separate and non-coherent address spaces, with interactions realized by MMIO and DMA. MMIO exposes the base address registers (BARs) of a peripheral XPU device in the host address space, enabling the CPU to launch uncached loads and stores to the device BAR space. However, PCIe’s high per-transaction latency and strict write-ordering, which allows only one outstanding write, limit the MMIO performance \cite{intel_accel_ecosystem}. On the XPU side, its DMA engine allows to directly access the host memory and delivers a much higher throughput for bulky contiguous transfers. But, DMA operations incur substantial per-transfer setup overhead, making fine-grained and randomly distributed data movements highly inefficient~\cite{demystifying_type2,understanding_pcie}.

\subsubsection{Collaborative computing mode}
Conventional heterogeneous computing units employ a descriptor-driven producer-consumer model in which the CPU takes charge of the control plane for task dispatch, while the XPU serves as the computing plane for accelerated computing. This mode is used by today’s mainstream XPUs, including GPUs, SmartNICs, and FPGAs, and its core process comprises a four-stage pipeline. Stage 1: The CPU allocates a DMA-safe memory region (e.g., pinned memory), copies the input data there, encapsulates the task in a structured descriptor and enqueues it, then rings a doorbell register via MMIO to notify the XPU of a new task. 
Stage 2: The XPU polls the doorbell and invokes its DMA engine to retrieve the pending descriptor from the host memory. 
Stage 3: Guided by the descriptor, the XPU issues DMA reads to fetch the input data, then performs the specified computation task. 
Stage 4: Upon completion, the XPU writes the results back to the host memory via DMA and then raises an event to notify the CPU of the results.
Although this decoupled approach maximizes throughput for transferring bulky contiguous data, the isolated address spaces on the CPU and XPU inevitably incur unnecessary data copies and hinder efficient sharing of complex data structures between compute units, thus significantly limiting collaborative computing efficiency.

\subsection{Compute Express Link}

%In recent years, several cache-coherent interconnect industry standards have emerged to overcome coordination inefficiencies and programming complexity in PCIe-based heterogeneous systems. In proprietary ecosystems, NVIDIA’s NVLink-C2C achieves up to \SI{450}{GB/s} unidirectional bandwidth between an ARM CPU and an H100 GPU, leveraging the AMBA CHI protocol for hardware-level cache coherence, which allows peer-to-peer access at cacheline granularity \cite{nvlink_c2c,nvidia_gh100_whitepaper}. Similarly, AMD’s Infinity Fabric within the MI300A APU employs Infinity Cache to provide coherent shared access to \SI{128}{GB} HBM3\cite{infinity_fabric}. Intel’s Xeon® scalable processors integrate a suite of on-chip accelerators, such as the data streaming accelerator (DSA) and advanced matrix extensions (AMX). These accelerators connect to the on-chip network via a coherent mesh interface, allowing them to fully take advantage of the powerful memory subsystem of the CPU~\cite{intel_accel_ecosystem,4th_Xeon}. Although these vendor-specific solutions deliver high performance, their cross-platform portability is very limited. In the open-standard domain, early contenders such as OpenCAPI~\cite{opencapi}, Gen-Z \cite{genz} and CCIX \cite{ccix} failed to gain widespread adoption due to compatibility, scalability, or performance constraints. Currently, all of these proposals have been integrated into the CXL~\cite{CXL_spec}.

Built on the PCIe physical layer, CXL comprises three sub-protocols: CXL.io provides PCIe-equivalent features; CXL.cache enables devices to coherently access host memory; CXL.mem allows the host to perform load/store operations directly on device-attached memory. By combining these sub-protocols, three device types are distinguished. Type-1 devices implement CXL.io and CXL.cache (e.g., SmartNICs without device memory). Type-2 devices support all three sub-protocols (e.g., GPUs with device memory and specialized compute units). Type-3 devices use CXL.io and CXL.mem as memory expanders. 

This paper focuses on type-1/2 devices, both of which can be recognized as accelerators (XPUs). Each device includes a \textit{host memory cache} (HMC) that caches host memory and acts as a peer to the CPU’s L2 cache, with the CPU's last-level cache (LLC) serving as a coherence synchronization point. The host tracks coherence among peer caches, while the CXL device interacts with the host over the \textit{device coherency engine} (DCOH) using a lightweight MESI protocol. In addition to standard load/store operations, CXL.cache introduces the \textit{non-cacheable push} (NC-P) instruction \cite{cxl_fpga_ip}, which allows a device to push a cacheline directly into the host LLC and invalidate it in the HMC, providing a low-latency path for returning computed results. Sections \ref{Sec:RPC} will demonstrate how NC-P optimizes acceleration to boost application performance.

% =====================================================
\section{Coherent Heterogeneous Computing Framework}
\label{Sec:Coherent_Heterogeneous_Computing_Framework}
% =====================================================

\subsection{Limitations of Conventional Heterogeneous Computing}
\label{Sec:Limitations_of_Conventional_Heterogeneous_Computing}
Traditional PCIe-based heterogeneous systems are becoming increasingly crippled in dealing with modern workloads such as LLM training/inference and big data analytics. We identify four fundamental limitations as follows.

\textbf{L1: Compute-memory coupling leads to resource under-utilization.} Current compute units, regardless of CPUs or XPUs, are tightly coupled to their own memory, as illustrated with the dotted boxes in Fig.~\ref{Fig:PCIe_based}. 
These host servers and accelerators form basic performance scaling units, leading to either compute or memory over-provisions in today's hyperscale datacenters \cite{borg,pond}.
%In addition, the distributed physical memories remain mutually isolated, without memory-semantic access and hardware coherence. 
%Mandatory data replication prevents the direct sharing of complex data structures and leads to poor resource utilization.
%This unfortunately results in mandatory data replications and infeasibility of direct sharing of complex data structures in heterogeneous computing, which sacrifice both computing efficiency and memory utilization. 
In addition, the distributed physical memories remain mutually isolated,
resulting in mandatory data copies and infeasibility of direct sharing of complex data structures in heterogeneous computing.

\textbf{L2: Inefficient collaborative mode.} The four-stage interaction model is optimized for processing bulky and contiguous data streams, but it incurs severe latency penalties under frequent fine-grained or random access patterns. This forces heterogeneous computing units to operate in isolation, preventing efficient collaboration in accomplishing complicated computation tasks.

\textbf{L3: CPU suffers from accelerator tax.} The CPU-centric master-slave computing paradigm forces all control and synchronization through the CPU. The CPU overheads, also referred to as accelerator tax, in task dispatch and result collection may offset the raw acceleration gains.

\textbf{L4: Complex programming model.} The lack of hardware coherence results in complex programming models that depend on explicit/implicit memory management by programmers/software, further constrained by limited device memory capacity. Developers face a fundamental dilemma: manual data placement might yield high performance but impose significant programming efforts, whereas relying on software-managed unified memory mechanisms (e.g., CUDA unified memory (UM) \cite{cuda_um}) simplifies development at the risk of performance loss due to expensive page fault handling.

%\subsection{CXL-based Coherent Heterogeneous Computing}
\subsection{Cohet Design Philosophy}
\label{Sec:CXL-based_Coherent_Heterogeneous_Computing}

\begin{figure}[tb]
\centering
\includegraphics[width=0.34\textwidth]{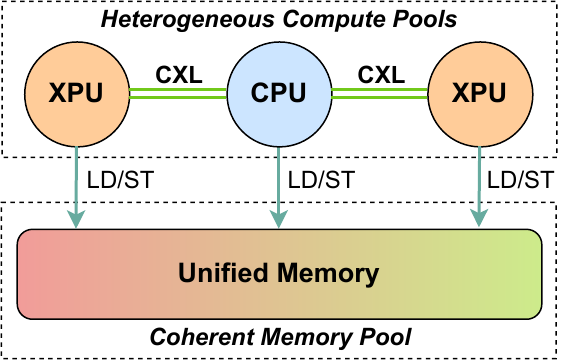}
\caption{Emerging CXL-based coherent heterogeneous computing architecture with unbiased cooperative computing and a coherent unified memory pool.}
\label{Fig:CXL_based}
\vspace{-15pt}
\end{figure}

To overcome the limitations of conventional heterogeneous computing and effectively leverage heterogeneous resources, CXL-based coherent heterogeneous computing offers a promising solution forward. 
%As illustrated in Fig.~\ref{Fig:CXL_based}, CPUs interconnect with various XPUs over CXL links, forming a memory-semantic fabric with hardware-maintained coherence. %Heterogeneous computing units access a unified memory pool via standard load/store instructions, enabling efficient data sharing and collaborative computing. 
%This architecture addresses the shortcomings of conventional heterogeneous systems in four key dimensions.
 Fig.~\ref{Fig:CXL_based} illustrates its architecture with the following four potential solutions.

\textbf{S1: Resource disaggregation \& pooling.} CXL decouples compute and memory resources, enabling the construction of heterogeneous compute and coherent memory pools. These disaggregated compute and memory resources can be expanded on demand, thus greatly boosting utilization. A unified memory view with hardware-maintained coherence helps CPUs and XPUs directly share large-scale data sets, avoiding unnecessary deep copies of working data and the associated software overhead.

\textbf{S2: Collaborative mode optimization.} CXL optimizes the interaction mode between CPUs and XPUs by providing fine-grained, low-latency memory-semantic accesses in both the CPU-to-XPU (CXL.mem) and XPU-to-CPU (CXL.cache) directions. Looking ahead, this capability sets the stage for the unification of two complementary communication mechanisms, with CXL.mem/cache handling latency-sensitive operations while CXL.io serving throughput-oriented transfers.

\textbf{S3: Decentralized control and synchronization.} 
%CXL enables decentralized synchronization by elevating XPUs to first-class compute agents. 
CXL promotes XPUs that are traditionally seen as subordinate devices to a peer position to CPUs. With access to coherent memory, CPUs and XPUs can now coordinate tasks and share data efficiently, without routing all controls to CPUs. Hardware-supported atomic operations across diverse devices replace traditional lock-based or software-coordinated synchronization, improving the performance of shared data workflows.

\textbf{S4: Programming model simplification.} Hardware-maintained coherence enables the OS to recognize heterogeneous compute units as standard NUMA nodes, providing unified memory allocation interfaces (e.g., malloc, mmap, C++ new), and allowing compute units to overcommit memory beyond the size of system memory. By simplifying heterogeneous programming to homogeneous while maintaining performance, this approach breaks through the developer dilemma in conventional heterogeneous computing. Moreover, existing applications can benefit seamlessly from high-performance heterogeneous acceleration without any refactoring or porting.

\begin{figure}[tb]
\centering
\includegraphics[width=0.48\textwidth]{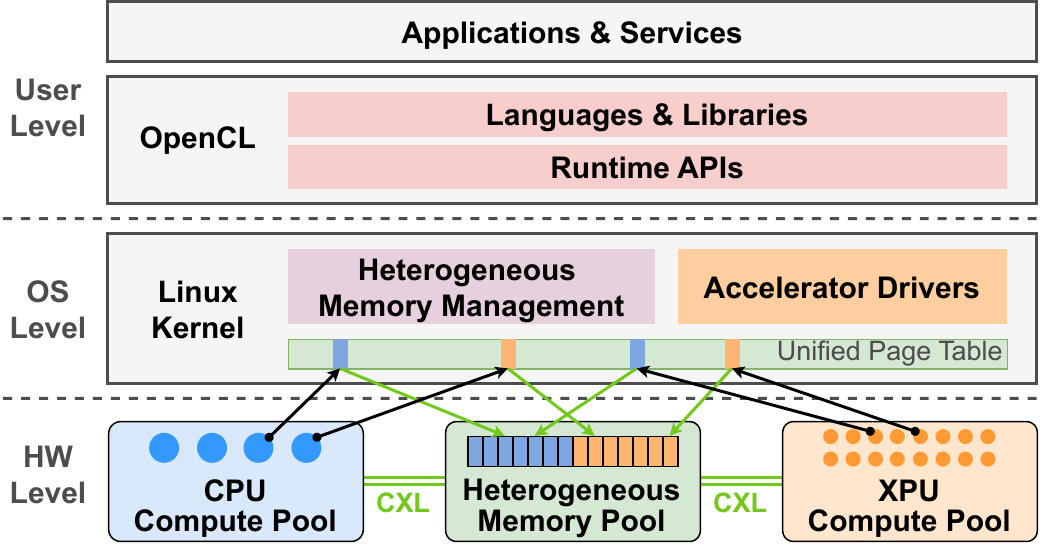}
\caption{Vertical stack of the Cohet computing framework.}
\label{Fig:Cohet_framework}
\vspace{-15pt}
\end{figure}

\subsection{Cohet Computing Framework}
In light of the above four guidelines, we propose Cohet, an open coherent heterogeneous computing framework. As shown in Fig.~\ref{Fig:Cohet_framework}, the vertical stack of Cohet is organized into three levels: hardware, operating system (OS), and user.

\subsubsection{Hardware level} One or more CXL switches compose a CXL fabric. A distributed resource scheduler (fabric manager) is implemented in each switch to allocate/release fabric-attached memory and XPU resources to a specific host. %The FM allocates/releases resources by binding/unbinding the logical ports of the CXL switches to hosts. 
For each host (OS point of view), its CPUs and XPUs interconnect via the CXL fabric to form a unified physically-addressed heterogeneous memory pool, presenting a single memory view to all compute units. CXL's hardware-maintained coherence not only allows CPUs and XPUs to cache and access each other’s memory at cacheline granularity but also supports cross-device atomic operations, enabling fine-grained and scalable synchronization across all running threads in the system. In this CXL-enabled architecture, the address translation service (ATS) lets CPUs and XPUs share a single per-process page table (denoted as Unified Page Table in the figure). When an XPU thread accesses a virtual address, it first looks up the mapping in its device-side address translation cache (ATC), analogous to the host TLB. Upon an ATC miss, the request is forwarded to the CPU-side IOMMU, which performs a page-table walk to resolve the physical address. The resolved mapping entry is then returned to the ATC, after which the XPU proceeds with its memory accesses \cite{CXL_spec}. At this level, tailoring the microarchitecture of endpoint devices (e.g., XPUs and memory expanders) for target applications is also crucial for maximizing the benefits of the coherent interconnect, as demonstrated with our CXL-NIC designs for accelerating two killer-apps in Section \ref{Sec:killer-apps}.

\subsubsection{OS level} The Linux kernel recognizes CPUs and XPUs as separate NUMA nodes. Heterogeneous memory management (HMM) merges device memory with host memory into the system memory pool, maintains the unified page table, and exposes a standard mmap/malloc interface to upper-layer software \cite{linux_hmm}. Specifically, the device driver probes the device memory size during system initialization, registers a device instance with HMM, and implements the callback functions required for unified page-table management, such as handling ATC invalidations and page migrations. Furthermore, we modify the kernel’s \textit{numa\_init} routine, which inspects the available system memory, initializes the host and device memory as distinct NUMA nodes based on their types, and binds them to the corresponding CPU or XPU. 

A malloc call allocates a page-table entry without assigning a physical frame, allowing memory overcommitment. On an XPU's first access to a given virtual address, an ATC miss triggers an IOMMU translation request. The kernel then updates the page-table entry to point to XPU physical memory. Once pages are allocated, both CPUs and XPUs access them at cacheline granularity under CXL’s coherence guarantee. By contrast, PCIe-based non-coherent interconnects incur high-overhead OS page-fault interrupts and DMA-based page copies, which prior studies identify as the primary performance bottleneck in frameworks such as UM \cite{um_overhead}. When the unified page table is about to be updated due to page migration or swapping, HMM invokes the registered driver callback. The driver then temporarily blocks the device from accessing the affected page-table entries, allowing HMM to safely perform the update and trigger the IOMMU invalidation process. According to the ATS protocol, this process invalidates the corresponding entries in the device-side ATC. Once the invalidation has been completed, HMM notifies the driver to resume device address translation. %In modern NUMA systems, highly optimized page-migration schemes (e.g., AutoNUMA~\cite{autonuma} and TPP \cite{maruf2023tpp}) place hot data close to compute. However, these traditional schemes may fail under the Cohet framework, because the ``temperature'' of a data page can differ between CPU and XPU access patterns and shared pages are at risk of thrashing. We leave the design of adaptive page migration schemes for Cohet systems to future work.
Moreover, adaptive page migration in Cohet is a potential performance optimization, left for future work.

\begin{figure*}[t]
\centering
\includegraphics[width=0.9\textwidth]{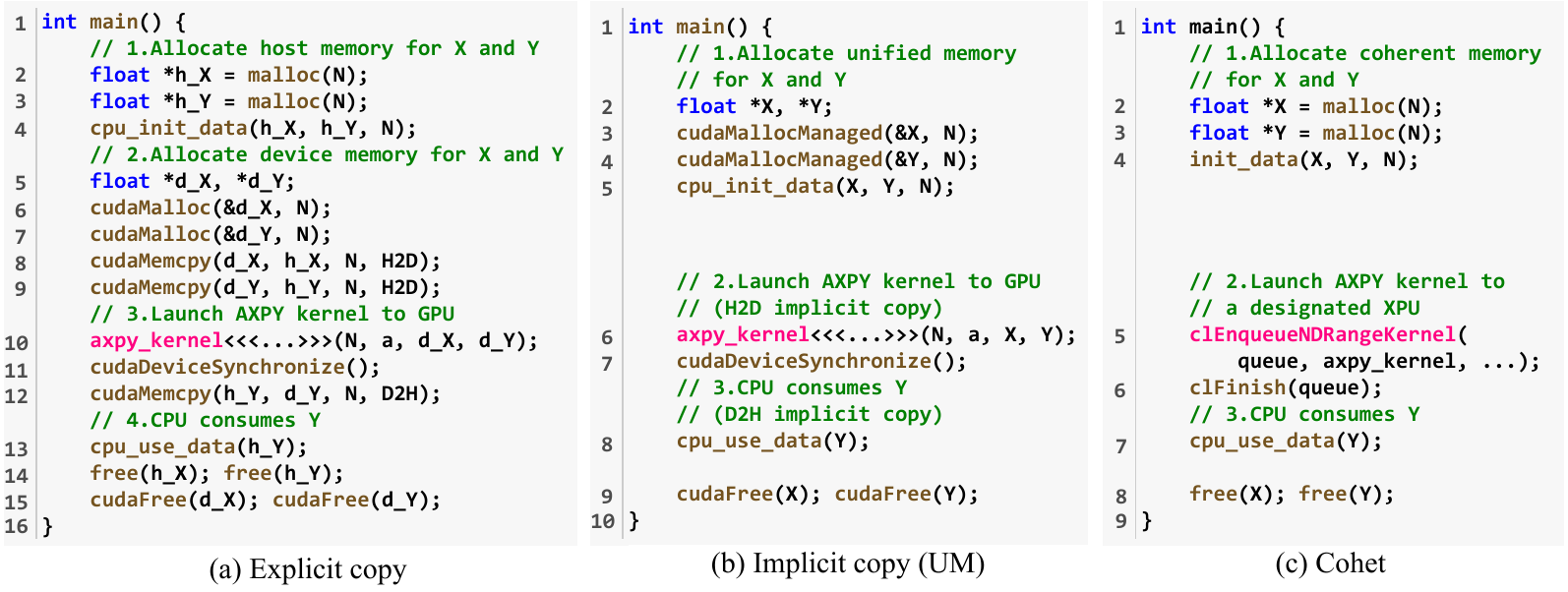}
\caption{Comparison between CUDA and Cohet programming models using AXPY operation as an example.} 
\label{Fig:code}
\vspace{-15pt}
\end{figure*}

\subsubsection{User level} Cohet embraces OpenCL 3.0 \cite{opencl}, the most pervasive cross-vendor open standard for low-level heterogeneous parallel programming, to deliver a fully transparent programming model. While the standard OpenCL requires special APIs for explicit (such as \textit{clHostMemAllocINTEL}) or implicit memory management (such as \textit{clSharedMemAllocINTEL}) before execution, Cohet allows applications to use mmap, malloc, and C++ new exactly as in CPU-only systems, without incurring page-fault penalties. Fig. \ref{Fig:code} compares Cohet with CUDA’s explicit and implicit copy programming models, using \textit{AXPY} operation ($\mathbf{Y} = \alpha \mathbf{X} + \mathbf{Y}$) as a pseudocode example. Fig. \ref{Fig:code}(a) relies heavily on programmers for manual memory management, resulting in the highest code complexity (16 lines). Fig. \ref{Fig:code}(b) employs UM to reduce programming effort (10 lines), but it incurs non-negligible overhead from implicit data copies. In contrast, Cohet in Fig. \ref{Fig:code}(c) further simplifies heterogeneous programming to a level approaching homogeneous programming (9 lines), while achieving performance comparable to manually managed data.

% =====================================================
\section{SimCXL Design and Implementation}
% =====================================================

\subsection{Simulator Architecture}
\label{Sec:Simulator_Architecture}
\begin{figure}[t]
\centering
\includegraphics[width=0.40\textwidth]{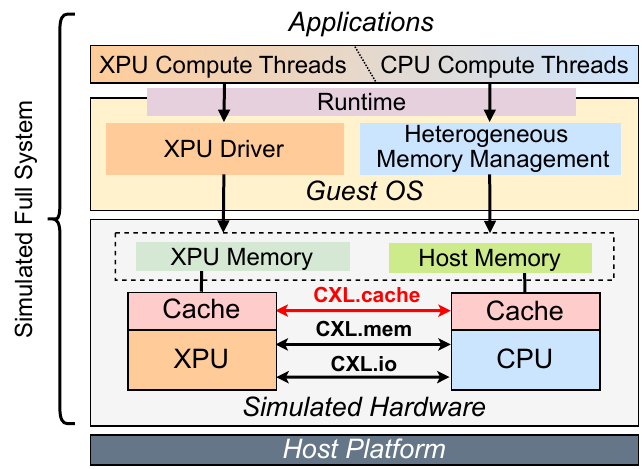}
\caption{Architecture of the proposed SimCXL simulator.} 
\label{Fig:SimCXL_overview}
\vspace{-16pt}
\end{figure}

SimCXL is a cycle-level simulator intended for modeling the proposed Cohet computing framework with a CXL interconnect fabric connecting CPUs, XPUs, and a memory pool in a cache-coherent manner. Fig.~\ref{Fig:SimCXL_overview} illustrates its modular architecture built on gem5's \textit{full-system} (FS) mode \cite{gem5_paper,gem5_v20}, modeling both the guest OS and hardware architecture to enable realistic simulation of software-hardware interactions. 

We build three main components on gem5 for SimCXL. First, we add a PCIe device model called XPU which comprises processing units, device-attached memory, and a local cache. Second, we implement the complete CXL sub-protocols: CXL.io supports device enumeration and configuration, register access, and DMA by extending gem5’s native PCIe protocol; CXL.cache and CXL.mem enable coherent load and store operations in device-to-host (D2H) and host-to-device (H2D) directions, respectively. These extensions unify host memory and device memory into a unified coherent memory pool through a custom MESI coherency protocol. Third, we develop a dedicated device driver in the guest OS to allow the host to discover and configure CXL devices; it also cooperates with the HMM module to present a unified mmap and malloc interface to user applications. %This enables a transparent and efficient collaborative computing paradigm, which allows user applications to flexibly map control-plane tasks or branch-heavy operations to the CPU compute units while offloading highly parallel or repetitive workloads to XPU compute units without explicit memory management.

\subsection{CXL Protocol Supports}
\label{Sec:CXL_protocol_support}

\begin{figure}
\centering
\includegraphics[width=0.47\textwidth]{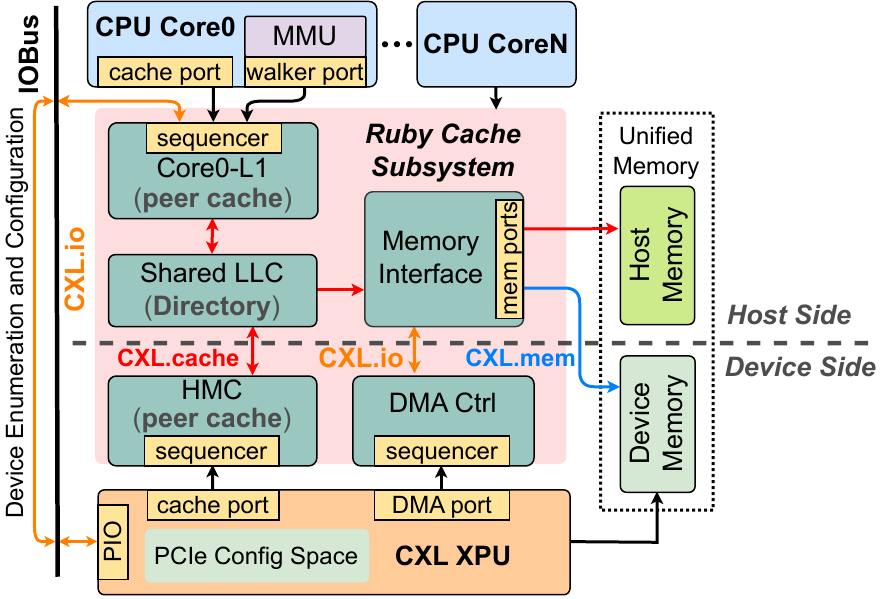}
\caption{CXL protocol implementation in SimCXL.}
\label{Fig:CXL_cache_proto}
\vspace{-10pt}
\end{figure}

Fig.~\ref{Fig:CXL_cache_proto} shows an x86 platform with a Ruby subsystem used in the FS mode. Ruby \cite{gem5_ruby} supports the flexible modeling of various cache coherence protocols by specifying state transitions of multi-level caches using the domain-specific language SLICC. On the host side, the CPU cores and their MMUs connect to the Ruby sequencer through cache and Walker ports, respectively.
The sequencer routes memory access requests by analyzing packet destination addresses: memory access requests are converted into \textit{RubyRequest} packets and forwarded to the cache hierarchy (red arrows), while IO accesses are directed to target devices via the IO bus (orange arrows). The memory interface coordinates requests from LLC and DMA controllers while converting \textit{RubyRequest} packets into gem5 memory packets for memory controllers. The device side includes the CXL accelerator with its HMC and device memory. The accelerator utilizes three CXL sub-protocols. CXL.io supports MMIO and DMA through PIO and DMA ports, respectively. CXL.cache enables coherent memory access in cacheline granularity via the cache port. CXL.mem incorporates device memory into the host memory address space, managed uniformly by the memory interface, allowing cores to perform standard load/store operations with full cache coherence.

\subsubsection{CXL.io support} The CXL.io sub-protocol handles device enumeration and configuration during system initialization. The BIOS performs CXL.io configuration reads to determine the size of each BAR register space, maps the corresponding physical address range, and writes the base addresses back via configuration writes. A kernel driver then creates \textit{/dev/cxl\_acc} and exposes open, mmap and release syscalls, allowing the CPU to read and write the BAR space of the CXL device via MMIO to control the device. The device also performs DMA operations through its DMA port.

\subsubsection{CXL.cache support} A cache port is implemented and connects to the HMC within Ruby for the CXL accelerator. Using SLICC, we developed a directory-based  two-level MESI protocol optimized for heterogeneous systems. Core0-L1 and HMC serve as peer caches which are privately owned by Core0 and the accelerator, respectively, and they both share the LLC. %Directory information is embedded in the metadata fields of the LLC cacheline.
The metadata of each LLC cacheline embeds directory information for coherence management, including a \textit{CacheState} field for transient or stable states, an \textit{ID} field (the unique identifier of each cache controller in Ruby) tracking the exclusive holder, and a \textit{bit vector} recording all sharers.

%Fig.~\ref{Fig:CXL_Prot_Flow} illustrates an example of  CXL.cache sub-protocol workflows when an accelerator exclusively modifies and writes back a dirty cacheline, comprising three phases as follows. \circledone~Read For Ownership: The accelerator requests exclusive access to a target cacheline within the HMC, which then issues a \textit{RdOwn} request to the LLC upon a cache miss. At this point, assume that the latest cacheline copy resides in CoreX-L1 with the modified (\textit{M}) state. When the LLC receives the request, it triggers a two-step operation: first, it sends a \textit{SnpInv} to invalidate the cacheline in CoreX-L1 (transitioning to the invalid (\textit{I}) state); second, it writes back the dirty data from CoreX-L1 to host memory and forwards it to the requesting HMC, where the corresponding cacheline transitions to the exclusive (\textit{E}) state. \circledtwo~Silent Modification: The accelerator with exclusive ownership can modify the cacheline locally without generating any coherence messages, upgrading the HMC's cacheline state to \textit{M}. \circledthree~Data Eviction: When evicting a dirty cacheline, the HMC issues a \textit{DirtyEvict} request to the LLC. After verifying the directory state, the LLC responds with a  globally observed writepull (\textit{GO-WritePull}) authorization, allowing write-back to host memory. Finally, the LLC issues a \textit{GO-I} command to the HMC, marking the corresponding cacheline as I state.

\begin{figure}
\centering
\includegraphics[width=0.5\textwidth]{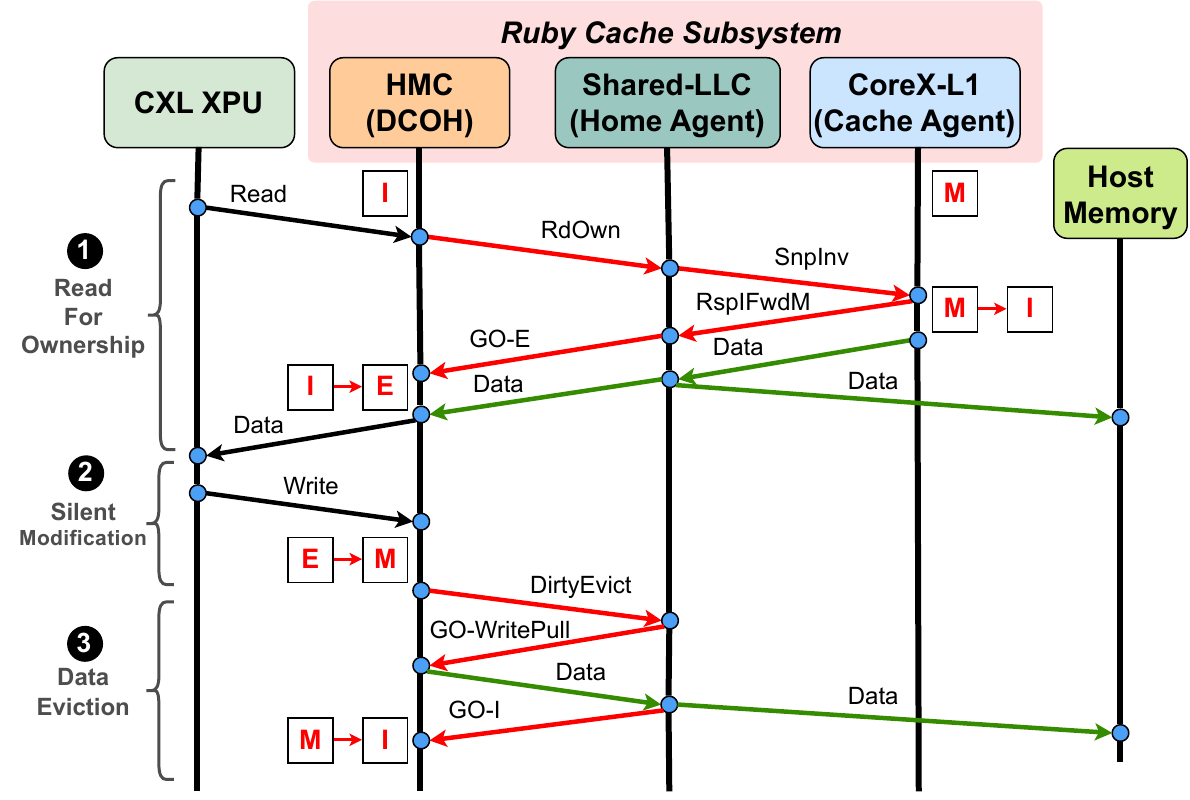}
\caption{An example of CXL.cache request-response flow when an XPU performs a store operation on a host memory address.}
\label{Fig:CXL_Prot_Flow}
\vspace{-15pt}
\end{figure}

Fig.~\ref{Fig:CXL_Prot_Flow} illustrates an example of  CXL.cache sub-protocol workflows when an accelerator exclusively modifies and writes back a dirty cacheline, in three phases. \circledone~Read For Ownership: The accelerator requests exclusive access to a target cacheline, which then issues a \textit{RdOwn} request to the LLC upon HMC miss. LLC sends \textit{SnpInv} to invalidate (\textit{I}) CoreX-L1's modified (\textit{M}) copy, writes back dirty data to memory, and forwards data with exclusive (\textit{E}) state to HMC. \circledtwo~Silent Modification: Accelerator modifies cacheline locally, upgrading state to \textit{M} without coherence messages. \circledthree~Data Eviction: HMC issues \textit{DirtyEvict}. LLC verifies state and authorizes write-back with \textit{GO-WritePull}, then sends \textit{GO-I} to invalidate HMC's cacheline.

\subsubsection{CXL.mem support}
%The CXL.mem sub-protocol integrates device memory into the host's physical address space, forming a unified physical memory view. It enables CPUs to access device memory directly and coherently just like local DDR memory. 
We developed a dedicated memory interface module for organizing the unified memory in our Ruby implementation. This module routes memory access requests from the shared LLC to either the host memory or the device memory based on address ranges configured by the BIOS and subsequently returns response data to the requester. The device memory can directly leverage various existing memory models in gem5, including DDR3/4/5, non-volatile memory (NVM), and high bandwidth memory (HBM). Furthermore, we modified the memory management module in the OS kernel to recognize CXL device memory as a CPU-less NUMA node during system initialization. This modification enables transparent utilization of CXL-expanded memory by upper-layer applications.

% =====================================================
\vspace{-3pt}
\section{Cohet Killer Apps}
\label{Sec:killer-apps}
% =====================================================
\subsection{Remote Atomic Operations}
\label{Sec:RAO}
Atomic operations play a pivotal role in parallel programming, as they are widely used to ensure consistency of shared data accesses \cite{dynamo} and implement lock-free algorithms~\cite{non_lock} as well as synchronization primitives such as spinlocks \cite{atomic_cache,lock_1,lock_2}. As a result, modern processors have been highly optimized for efficient atomic operations within a single multi-core node. However, the emergence of data-intensive workloads (e.g., machine learning and graph analytics) has driven the distribution of massive shared data sets across multiple computing nodes in order to exploit data parallel processing. This shift gives rise to the demand for frequent cross-node \textit{remote atomic operations} (RAOs)\cite{circustent,RAE,rao_sequencer,think_more_rdma,rdma_design,lock_2,ao_evaluating}. RAOs enforce atomic updates to shared data residing in a remote memory location, which is an extension of local atomics to distributed computing systems by means of remote direct memory access (RDMA). The study in~\cite{RAE} shows that when the number of computing nodes exceeds 64 and global memory accesses follow a random indexing pattern, RAOs can constitute as much as 98.44\% of all atomic operations. Hence, the design and implementation efficiency of RAOs are crucial to the performance of large-scale high-performance computing systems.

\begin{figure}[t]
\centering
\includegraphics[width=0.5\textwidth]{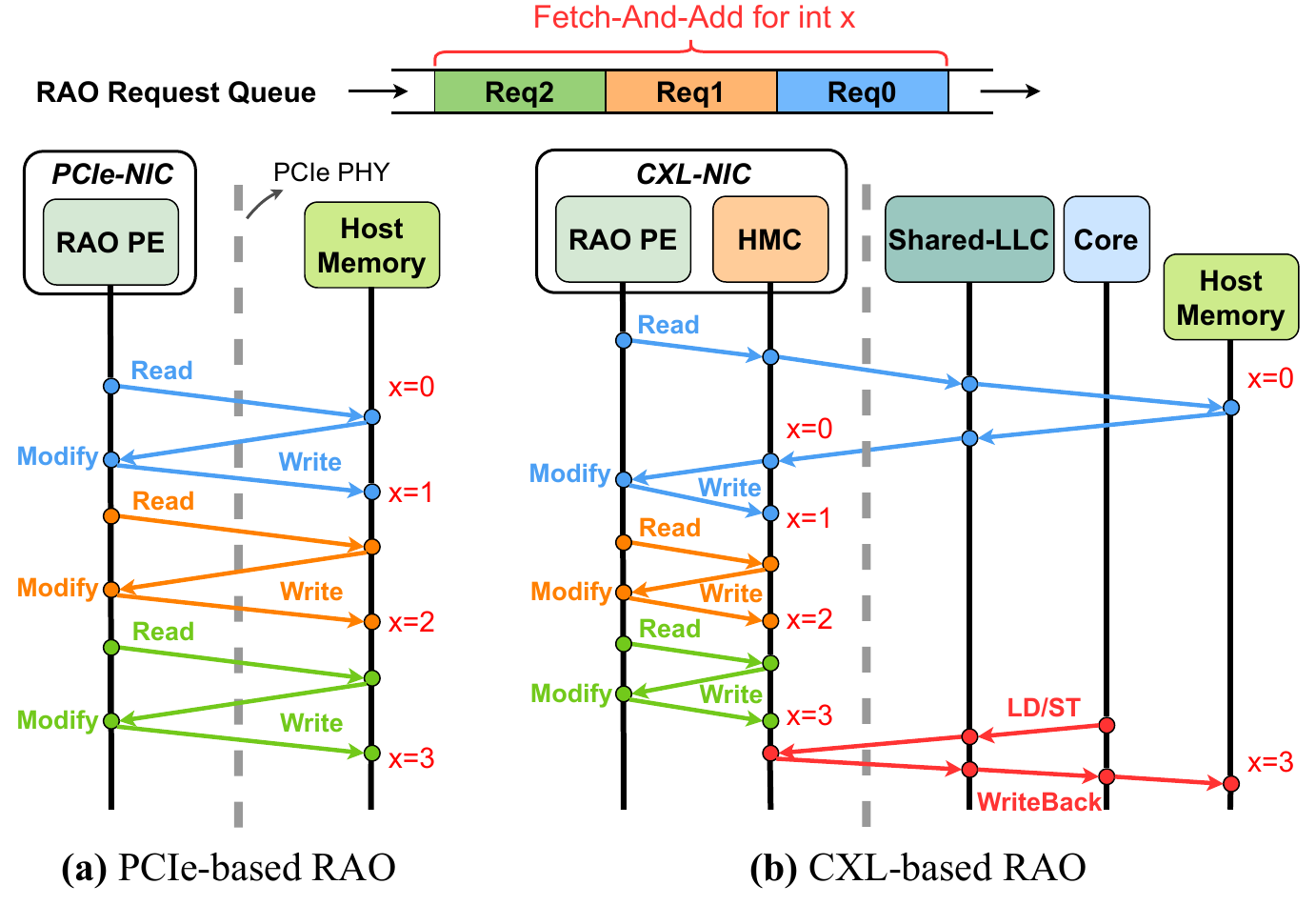}
\caption{PCIe-based RAO vs. CXL-based RAO in NIC design.}
\label{Fig:RAO_Comp}
\vspace{-15pt}
\end{figure}

\subsubsection{Conventional RAO offloading design in PCIe-NIC}
Traditional software-based implementations of RAOs rely on parallel programming libraries such as MPI and OpenSHMEM, but these implementations often suffer from performance overhead due to redundant software routines (e.g., multi-layer communication protocol stacks, software locks/semaphores, and redundant memory copies) \cite{circustent,RAE}. 
%To address this problem, modern supercomputers employ advanced interconnect protocols like remote direct memory access (RDMA) over InfiniBand \cite{infiniband}, which optimize common RAO primitives such as compare-and-swap (CAS) and fetch-and-add (FAA) through near-data processing techniques.
To address this problem, modern supercomputers typically employ hardware offloading techniques in RDMA networks~\cite{infiniband} to optimize common RAO primitives such as \textit{compare-and-swap} (CAS) and \textit{fetch-and-add} (FAA).
In such systems, RAOs are offloaded to the RDMA network interface controllers (NICs) and executed as indivisible \textit{read-modify-write} operations across the PCIe interface, significantly reducing inter-node traffics. However, PCIe-NICs still face critical limitations when accelerating RAOs. That is, PCIe does not intrinsically guarantee memory consistency and atomicity, as well as timely visibility of data modifications to the host. Each RAO requires two DMA transfers, one for the read and another for the write, which must be issued consecutively without interruption for the same target address. This can be explained with an example illustrated in Fig.~\ref{Fig:RAO_Comp}(a). When three FAA requests on a shared int variable $x$ are queued for processing, the NIC must launch six costly DMA transactions across PCIe. Moreover, due to PCIe’s relaxed ordering and split-transaction mechanisms \cite{introduction_cxl}, a later read request may arrive before a prior write, leading to potential read-after-write (RAW) hazards. To avoid such hazards, each RAO must wait for an acknowledgment of the previous write before proceeding, which significantly limits the RAO throughput~\cite{free_atomic}. 

%\subsubsection{Implementation}
\subsubsection{Cohet-driven RAO offloading design in CXL-NIC}

In contrast, Fig.~\ref{Fig:RAO_Comp}(b) shows the CXL-based RAO design guided by Cohet to address this bottleneck. The CXL-NIC and the host CPU share a unified memory view. This allows the CXL-NIC to cache the target variable in the HMC, enabling subsequent RAOs to be serviced directly within the HMC without redundant PCIe transfers. The processing element (PE) locks the target RAO cacheline to prevent any invalidation during the RMW operation and thus preserve atomicity. Meanwhile, hardware-maintained cache coherence ensures that the host always perceives and retrieves the most recent data. This mechanism substantially improves the efficiency of RAO-intensive operations such as sequencers \cite{rao_sequencer}, lock services \cite{lock_1,lock_2}, and synchronization barriers \cite{atomic_cache}, which often involve many-to-one contention on a small set of hot memory locations. Furthermore, most data-intensive applications use pointer-based data structures such as graphs, unbalanced trees, unstructured grids, and sparse matrices. These structures generate fine-grained and irregular memory access patterns \cite{graphpim}. As discussed in Section \ref{Sec:Conventional_Heterogeneous_Computing}, traditional DMA mechanisms are unsuitable for such access patterns. In contrast, CXL.cache efficiently supports such workloads through low-latency, coherent access to shared memory.

\begin{figure}[tb]
\centering
\includegraphics[width=0.43\textwidth]{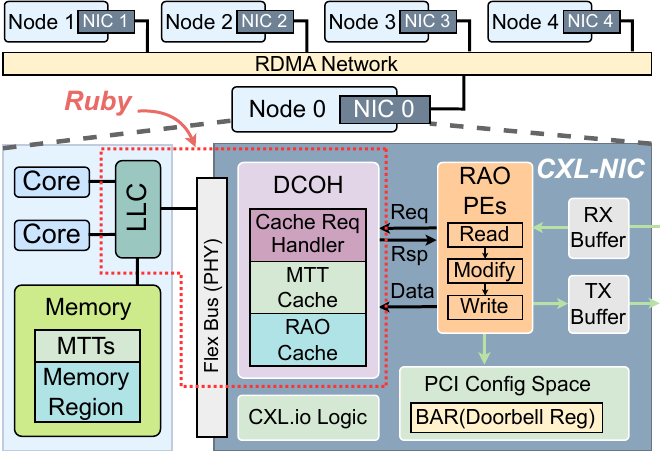}
\caption{RAO offloading design in a CXL-NIC of SimCXL.}
\label{Fig:CXL_RAO}
\vspace{-15pt}
\end{figure}

As shown in Fig.~\ref{Fig:CXL_RAO}, we implemented a CXL-NIC with RAO offloading design in SimCXL. To boost concurrency, multiple RAO PEs are added to handle incoming RAO requests from remote servers in parallel. The DCOH module is responsible for caching data and maintaining coherence with the host’s cache hierarchy. When an RAO request arrives at the RX buffer, the RAO PEs parse the request and perform the read-modify-write operation in three stages. (1) In the ``read'' stage, a PE first queries the DCOH to check whether the target variable resides in the cache. In case of a cache hit, the operation proceeds directly to the ``modify'' stage. Otherwise, the DCOH requests the latest data from the host’s LLC. (2) In the ``modify'' stage, the PE executes the atomic operation according to the request type, such as FAA or CAS. (3) In the ``write'' stage, the result is written back to the cache without immediate eviction to the host. When the host later accesses the same variable, the DCOH ensures that any dirty cachelines are written back to host memory (as shown in Fig.~\ref{Fig:RAO_Comp}). After the RAO operation completes, the CXL-NIC sends the response back to the remote server.

\subsection{Remote Procedure Calls}
\label{Sec:RPC}
Remote procedure calls (RPCs) serve as the fundamental inter-process communication layer for modern cloud services, supporting diverse distributed applications (e.g., microservices~\cite{micro_server}, cloud storage \cite{cloud_storage}, and machine learning \cite{ML}). An RPC abstracts remote compute resources by making a function call to a remote machine similar to a local call. Its underlying stack (typically implemented as a user-space library \cite{grpc,thrift}) transparently handles connection management, network protocols, parameter (de)serialization, and thread scheduling. 
%The (de)serialization process, despite enabling applications to easily exchange data in an architecture- and language-independent manner, represents a high-cost operation. 
Industry measurements reveal significant RPC overheads: Google reports that RPCs account for 7.1\% of CPU cycles in production clusters \cite{google_2023}; Meta observes that parameter (de)serialization alone consumes 6.7\% of CPU cycles in seven critical microservices \cite{meta_2020}. These findings highlight the importance of accelerating the RPC stack especially (de)serialization to reclaim CPU resources for application workloads.

%\subsubsection{Workflow analysis}
\subsubsection{Conventional RPC offloading design in PCIe-NIC}
Among existing serialization libraries \cite{noauthor_protocol_nodate,noauthor_capn_nodate,noauthor_flatbuffers_nodate}, we focus on Google’s Protobuf \cite{noauthor_protocol_nodate} in this paper, as it has widespread adoption in today's cloud applications. 
%(we argue that CXL NIC can similarly improve offload performance for newer libraries such as Cap’n Proto \cite{noauthor_capn_nodate} and FlatBuffers~\cite{noauthor_flatbuffers_nodate}). 
Two key characteristics of RPC messages critically impact acceleration designs. (1) The vast majority of messages are extremely small; a Google cluster analysis shows that 56\% of messages are $\leq$$32$ bytes and 93\% are $\leq$$512$ bytes \cite{protoAcc}. (2) Nested messages can be defined via pointer references, analogous to pointer chasing, incurring significant cumulative overhead during (de)serialization\cite{micro_server}; in real-world applications, the nesting of RPC message may exceed ten levels~\cite{protoAcc,google_2023}. %Prior work offloaded the RPC stack to on-chip accelerators~\cite{protoAcc,optimus_prime,on_chip3} or used customized interconnects \cite{dagger} to leverage low-latency memory accesses. However, these solutions require intrusive CPU modifications and incur prohibitive hardware development/deployment costs, impeding their adoption in commercial servers.

\begin{figure}[t]
\centering
\includegraphics[width=0.45\textwidth]{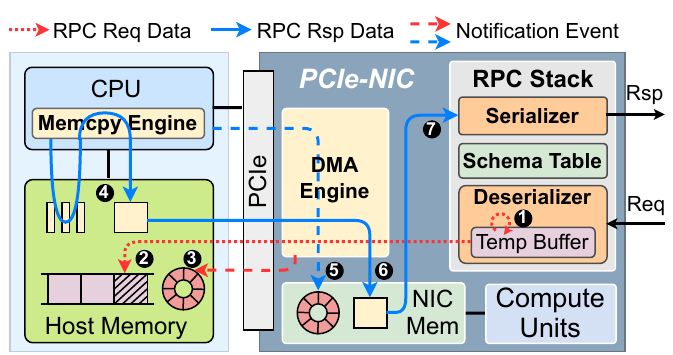}
\caption{Implementation of an RPC offloading design \cite{rpcnic} in a PCIe-NIC of SimCXL.}
\label{Fig:PCIe_RPC}
\vspace{-15pt}
\end{figure}

A recent study RpcNIC \cite{rpcnic} proposed a hardware offloading design of the RPC stack to a PCIe-NIC, which we have implemented in SimCXL as shown in Fig.~\ref{Fig:PCIe_RPC}.
%illustrates the acceleration process for de/serialization on a PCIe NIC. 
The host pre-runs the Protobuf compiler to store message structure metadata in a schema table, which guides message fields to decode in in-memory C++ objects or encode them into binary sequences. \circledone~When an RPC request arrives, the NIC deserializer decodes it field-by-field, accumulating results in a \SI{4}{KB} on-chip temp buffer. \circledtwo~When  the current request's deserialization is finished or the buffer is full, a one-shot DMA transfer to host memory is triggered. \circledthree~The NIC then increments the head pointer of a ring buffer in main memory through a DMA write to signal the arrival of a new request; the CPU processes the request and increments the tail pointer to indicate completion. \circledfour~For RPC responses, RpcNIC employs a pre-serialization mechanism; the CPU first invokes the on-chip memory copy engine (i.e., Intel's DSA) to iteratively copy noncontiguous fields into a pre-allocated DMA-safe buffer. \circledfive~The CPU then updates an NIC-resident ring buffer via MMIO to signal the completion of pre-serialization. \circledsix~The NIC copies the prepared data from the host memory via a DMA read. \circledseven~The hardware serializer iteratively encodes the data in the NIC memory and transmits the RPC response over the network. 

\begin{figure}[t]
\centering
\includegraphics[width=0.45\textwidth]{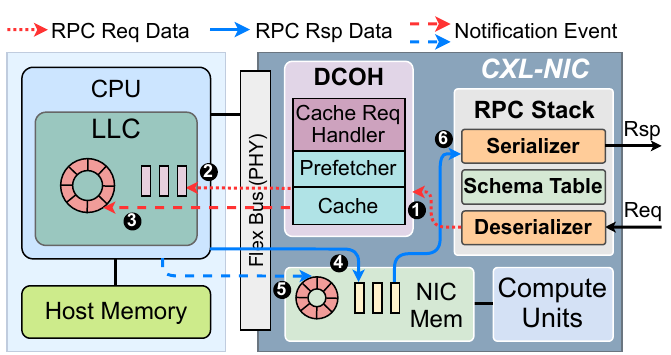}
\caption{RPC offloading design in a CXL-NIC of SimCXL.}
\label{Fig:CXL_RPC}
\vspace{-15pt}
\end{figure}

Despite the one-shot DMA and pre-serialization mechanism in the RpcNIC design mitigate the high latency and low throughput of fine-grained or nested RPC message transfers over PCIe, three inherent limitations remain: (1) high design complexity in multi-device interaction, (2) redundant data copies, and (3) non-trivial CPU control overhead that may negate acceleration benefits. %We argue that CXL-based NIC can fundamentally simplify hardware offloading by enabling zero-copy (de)serialization with truly CPU-transparent acceleration.

\subsubsection{Cohet-driven RPC offloading design in CXL-NIC}
Fig.~\ref{Fig:CXL_RPC} illustrates our proposed CXL-based NIC offloading design, where the NIC operates as a type-2 device that enables both the CPU and the NIC to access each other's memory through load/store instructions. The RPC processing pipeline comprises six phases as follows. \circledone~When an RPC request arrives, the CXL-NIC deserializer performs field-by-field decoding. \circledtwo~Each decoded field when ready is pushed by DCOH to a designated location in the CPU's LLC via NC-P (since the CPU will soon access the decoded fields to service the RPC request). \circledthree~A ring buffer is maintained in host memory and, more importantly, cached in the LLC for task queuing. \circledfour~For RPC responses, the CPU constructs and populates message objects to device memory using CXL.mem. \circledfive~Once the objects are ready, the CPU notifies the NIC of pending serialization tasks. \circledsix~The serializer performs iterative encoding from the NIC memory and transmits the RPC response over the network.

Although CXL.mem-based message construction delivers high performance, requiring application code modifications poses backward compatibility challenges for legacy systems. We also propose an alternative CXL.cache approach: The CPU constructs response messages in the host memory as done in conventional implementations, while the NIC uses CXL.cache to directly access these structures during serialization. 
To further reduce access latency, we enhance the DCOH module with a device-side hardware prefetcher, thanks to the coherenet memory-semantic interconnect.
%The coherent interconnect enables integration of device-side hardware prefetchers, so we implemented a multi-stride prefetcher in the DCOH module to further reduce access latency. 
The RPC prefetcher is a multi-stride prefetcher, which records cache-miss addresses to identify data streams with various stride patterns and issues prefetches accordingly, achieving a balance between performance and design complexity.
%To reduce access latency, we enhance the DCOH module with a device-side hardware prefetcher, which tracks memory access patterns and issues prefetches when \mbox{appropriate}.

\section{Experiments and Evaluation}
\label{Sec:Experiments_and_Evaluation}
\subsection{Experimental Setup}
\label{Sec:Experimental_Setup}

\begin{scriptsize}
\begin{table}[t]
\centering
\renewcommand{\arraystretch}{1.0} 
\caption{Configurations for hardware testbed and SimCXL.}
\label{tab:experimental_setup}
\resizebox{0.5\textwidth}{!}{
\begin{tabular}{|l|l|l|}
\hline
\textbf{Config. Parameter}  & \textbf{CXL Testbed} & \textbf{SimCXL}\\ 
\hline \hline
{Linux kernel version} & {v6.5.0} & {Modified v6.12} \\ \hline
{CPU type} & {Xeon® Platinum 8468V } & {X86O3CPU} \\ \hline
{CPU cores} &{48} & {48} \\ \hline
{Local DRAM type} & {DDR5 4800} & {DDR5 4400} \\ \hline
{\#Memory channels/NUMA} & {2} & {2} \\ \hline
{DDR DRAM size} & {1TB} & {32GB}   \\ \hline
{LLC size} & {97.5MB} & {96MB} \\ \hline
{CXL\&PCIe accelerators} & {Intel Agilex I-Series FPGA} & {CXL-\&PCIe-NIC models} \\ \hline
{HMC size} & {128KB, 4 ways} & {128KB, 4 ways} \\ \hline
{CXL memory expander} & {Samsung memory expander} & {Memory expander model} \\ \hline
\end{tabular}
}
\vspace{-10pt}
\end{table}
\end{scriptsize}

\subsubsection{Hardware testbed} 
Our hardware testbed is a high-performance dual-socket server, detailed in Table \ref{tab:experimental_setup}. Each socket houses an Intel Gen4 Xeon Platinum 8468V processor, which integrates four CPU chiplets with each paired with a memory controller driving two memory channels. Each channel is populated with two \SI{32}{GB} DDR5 \SI{4800}{MT/s} DIMMs. The four chiplets can be configured as a monolithic CPU or two/four NUMA nodes in sub-NUMA cluster mode (SNC)~\cite{4th_Xeon}. To ensure accurate and stable D2H memory access measurements, we enabled SNC-4 in the BIOS, dividing each socket into four separate NUMA nodes. Memory accesses from the same physical core to different NUMA nodes incur varying latencies depending on the number of NoC and UPI routing hops, thus manifesting the NUMA effect.
The CXL devices on the testbed include an Intel Agilex I-Series FPGA Development Kit \cite{agilex} and a Samsung \SI{512}{GB} memory expander \cite{samsung_cmm_d}. The FPGA operates at \SI{400}{MHz} and supports both CXL type-1/2 device configurations (denoted as CXL-FPGA) and standard PCIe configurations (denoted as PCIe-FPGA),  connected to Socket1 via PCIe5.0 ×16. The CXL-FPGA integrates a hard CXL IP \cite{cxl_fpga_ip} and is equipped with a \SI{128}{KB} 4-way HMC, whereas the PCIe-FPGA integrates a multi-channel DMA IP \cite{multichannel_dma}. %Unless otherwise specified, both CXL-FPGA and PCIe-FPGA default to accessing memory allocated on the nearest NUMA node (node 7). 
The Samsung expander, connected to Socket1 via PCIe5.0 ×8, is exposed to the system as a CPU-less NUMA node.

\subsubsection{SimCXL simulator}
SimCXL is developed  based on gem5 v23.1, with X86O3CPU and \SI{32}{GB} of \SI{4400}{MT/s} DDR5 dual-channel memory. In terms of CXL devices, the device driver is developed on top of the Linux v6.12 kernel. The modular CXL controller in SimCXL can model CXL type-1/2/3 devices. Device memory can directly leverage gem5’s native DDR, NVM, and HBM models; device compute units are customized following the gem5 programming framework. For type-1/2 devices (denoted as CXL-FPGA\textsubscript{sim}), a 4-way \SI{128}{KB} HMC is implemented in the Ruby subsystem. Based on the latency breakdowns presented in the CXL specification~\cite{CXL_spec} and related studies \cite{introduction_cxl,demystifying_type2,demystifying_type3,exploring_cxl_asic}, the CXL controller and Ruby’s heterogeneous MESI protocol set multiple configurable parameters for tuning H2D and D2H access latency and bandwidth. For PCIe devices (denoted as PCIe-FPGA\textsubscript{sim}), we implemented DMA read and write engines to handle H2D and D2H data transfers, respectively. 
%We used SimCXL at \SI{400}{MHz} to simulate hardware testbed configurations for CXL-FPGA and PCIe-FPGA and, simultaneously, simulated CXL-ASIC and PCIe-ASIC devices at \SI{1.5}{GHz} by frequency scaling the cycle counts obtained from the CXL-FPGA tests. This methodology enables us to evaluate the offloading performance of RAO and RPC on future production-grade CXL devices.
Both the CXL and PCIe device models can be configured to \SI{400}{MHz} to match the FPGA configurations. Meanwhile, the frequency can also be set to \SI{1.5}{GHz} to model real-world ASIC devices (denoted as CXL-ASIC\textsubscript{sim} and PCIe-ASIC\textsubscript{sim}, respectively), by frequency scaling based on the clock cycles measured from CXL-FPGA tests; this allows us to evaluate the realistic offloading performance of RAO and RPC on future production-grade CXL devices.

\subsubsection{Hardware calibration microbenchmarks}
To characterize the performance of CXL.cache, we implemented a load/store unit (LSU) on the CXL-FPGA and in SimCXL to generate host memory requests with configurable access patterns. On the FPGA, a purpose-designed performance monitoring unit collects latency and bandwidth statistics by recording request and response timestamps. In SimCXL, the device model includes detailed statistics tracking for the same measurements.
For DMA performance, we utilized the DMA IP on the PCIe-FPGA and the DMA read/write engine in SimCXL to generate workloads with varying message granularities and measured latency and bandwidth.

\begin{figure}[t]
\centering
\includegraphics[width=0.48\textwidth]{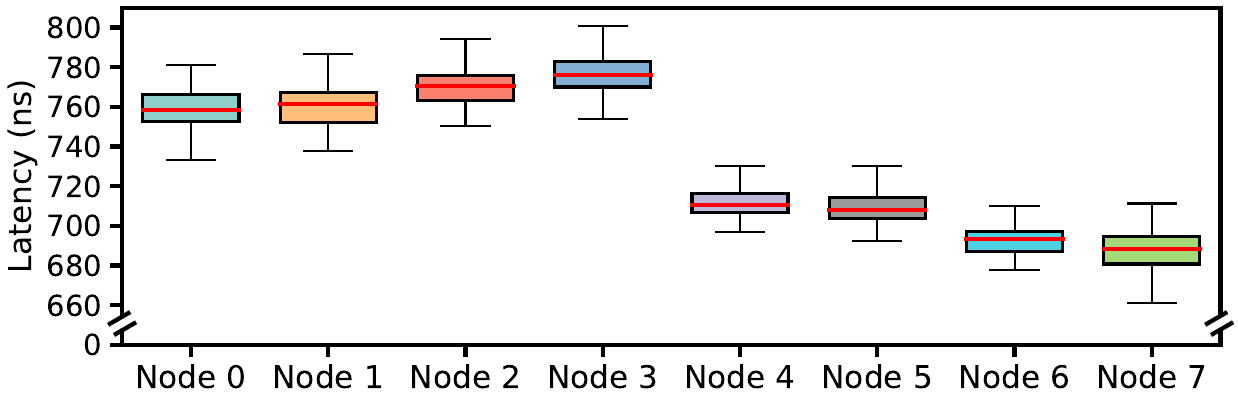}
\caption{Latency distribution of cacheline loads by the LSU via CXL.cache from memories on different NUMA nodes.} 
\label{Fig:numa_latency}
\vspace{-15pt}
\end{figure}

\subsubsection{Hardware calibration methodology}
In this work, we calibrate the latency and bandwidth of CXL.cache for D2H memory accesses. The calibration of H2D memory access via CXL.mem will be presented in another paper. %A load/store unit (LSU) is designed on the CXL-FPGA and SimCXL to generate desired host memory requests. The latency and bandwidth statistics on FPGA are collected by a purposely-designed performance monitoring unit which records request and response timestamps.SimCXL's device model includes detailed statistics tracking for accurate latency and bandwidth measurement. 
By tuning the model parameters of SimCXL, we are able to calibrate it to make its performance match that of CXL-FPGA. Note that the DMA performance calibration for SimCXL and PCIe-FPGA follows a similar approach. For D2H accesses via CXL.cache, cachelines may hit in the device's HMC, host LLC, or host memory. HMC hits are tested by repeating address sequences. For LLC hits and memory hits, after initializing the target host memory addresses, we use the \textit{CLDEMOTE} \cite{cldemote} instruction to push cachelines to the LLC or the \textit{CLFLUSH} \cite{clflush} instruction to flush cachelines to memory, followed by the LSU test sequence. For predictable performance, hyper-threading and hardware prefetchers are disabled, and CPU frequency is fixed at 2.4GHz during test.

% \subsubsection{Benchmark}
% To assess the performance gains of the CXL-based RAO and RPC, we used CircusTent atomic system benchmarks \cite{circustent} and Google HyperProtoBench \cite{protoAcc}, respectively. CircusTent implements varying atomic memory access patterns using FAA and CAS. HyperProtoBench comprises six benchmarks for Protocol Buffers (de)serialization, each processing 10 representative Google service messages.

\subsection{SimCXL Latency Calibration}
\label{Sec:SimCXL_Latency_Calibration}

\subsubsection{NUMA effects for CXL.cache}
We first evaluated the impact of NUMA effects on CXL.cache accesses by conducting identical tests across nodes 0-7 in sequence: we allocated and initialized huge pages on a target node, followed by LSU issuing 32 cacheline granularity (\SI{64}{B}) load requests to that node’s memory region. Results from 1000 repeated trials are shown in Fig.~\ref{Fig:numa_latency}. The CXL-FPGA accesses to the nearest node (node 7) exhibit the lowest median latency (\SI{688}{ns}). Within the same socket, the median latency increases with node distance: node 6 (\SI{693}{ns}), node 5 (\SI{708}{ns}), and node 4 (\SI{710}{ns}). In contrast, accesses to remote socket nodes incur substantially higher median latencies: node 0 (\SI{758}{ns}), node 1 (\SI{761}{ns}), node 2 (\SI{770}{ns}), and node 3 (\SI{776}{ns}). The median latency difference between the farthest node 3 and the nearest node 7 is \SI{88}{ns}, with a maximum latency gap approaching \SI{150}{ns}.
These results demonstrate a significant impact of NUMA effects on CXL.cache access latency, which suggests that the default memory allocation strategy (SNC-disabled) may scatter pages randomly across system memory, leading to unpredictable and often suboptimal access performance. Despite these striking latency disparities, the bandwidth measurements across nodes 0-7 show marginal variations. 

\begin{figure}[t]
\centering
\includegraphics[width=0.49\textwidth]{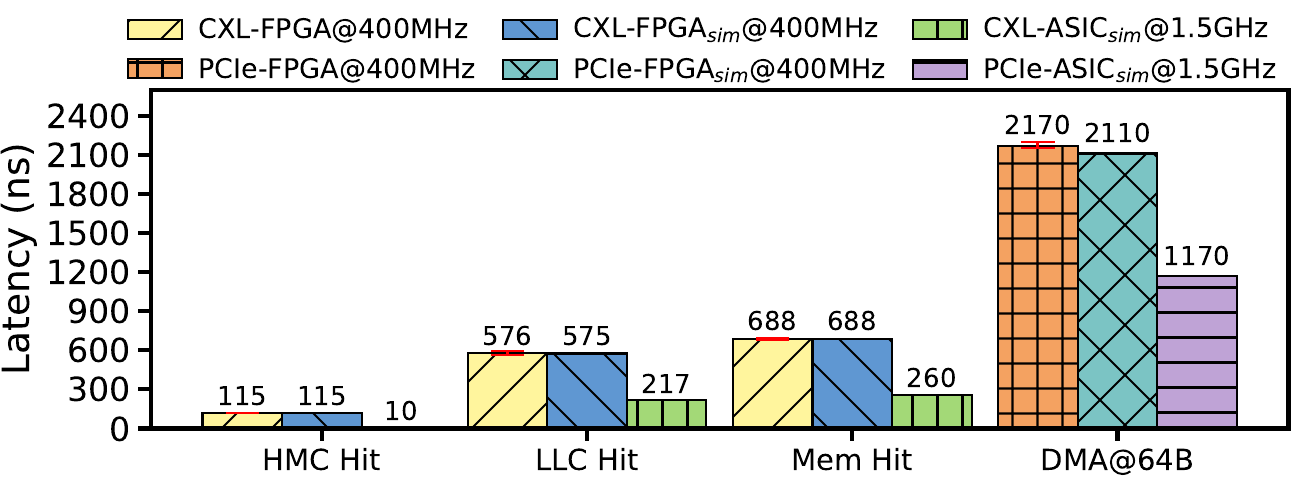}
\caption{Median load latency for cacheline HMC hits, LLC hits, and memory hits via CXL.cache, in comparison to DMA read latency at a \SI{64}{B} message granularity. Error bars represent the measured 25th and 75th percentiles.}
\label{Fig:latency}
\vspace{-10pt}
\end{figure}

\begin{figure}[t]
\centering
\includegraphics[width=0.48\textwidth]{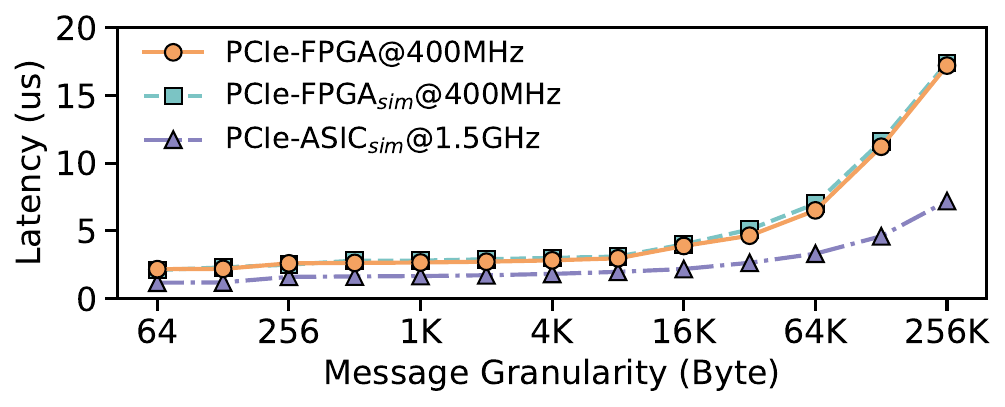}
\caption{Median H2D latency for DMA read with varying message granularity.} 
\label{Fig:dma_latency}
\vspace{-15pt}
\end{figure}

\subsubsection{CXL.cache latency}
To measure the latency of CXL.cache, we configured the LSU to issue 32 \SI{64}{B} load requests to sequential addresses, repeating each test 1000 times and reporting the median latency (only load results are shown, since PCIe PHY read/write performance is symmetric). As shown in Fig.~\ref{Fig:latency}, on the CXL-FPGA at \SI{400}{MHz}, HMC hit exhibits the lowest latency (\SI{115}{ns}), followed by LLC hit (\SI{575.6}{ns}), and off-chip memory hit (\SI{688.3}{ns}). The LLC hit latency is roughly 5$\times$ higher than the HMC hit, mainly due to the extra PCIe traversal after an HMC miss. Memory hit latency is only 1.19$\times$ higher than that of the LLC hit, reflecting a $\sim$\SI{100}{ns} DRAM access overhead. SimCXL results closely match the CXL-FPGA measurements, and their relatively high latencies come from the lower operating frequency at \SI{400}{MHz}. When we scale the frequency with the same clock cycle counts to \SI{1.5}{GHz} in SimCXL, CXL.cache shows the potential for much lower latencies. These results characterize the performance of CXL.cache across different memory access tiers, from HMC, through LLC over PCIe, to the most distant memory, and validate the fidelity and parameterized flexibility of SimCXL in modeling real-world CXL devices.

\subsubsection{DMA latency}
Fig. \ref{Fig:dma_latency} shows the latency curves for H2D DMA read transfers at different message granularities, for the purpose of comparing with the CXL.cache load. On a PCIe-FPGA operating at \SI{400}{MHz}, the latency remains roughly constant at $\sim$\SI{2.5}{us} for message sizes below \SI{8}{KB}, since the intrinsic DMA setup overhead is dominant at small  size. As the message size grows beyond \SI{8}{KB}, the data transfer time gradually dominates the overall latency. In contrast, CXL.cache incurs no setup overhead and performs memory accesses through load/store operations at cacheline granularity, achieving a 68\% lower latency for memory hit than DMA at 64B (see Fig. \ref{Fig:latency}). This result highlights the clear performance advantage of CXL.cache for fine-grained and latency-sensitive data accesses. Moreover, the DMA engine implemented in SimCXL at \SI{400}{MHz} closely matches the measured DMA latency curve on real platform. 
%We also extend it to the levels expected on an ASIC running at \SI{1.5}{GHz}.

\begin{figure}[t]
\centering
\includegraphics[width=0.49\textwidth]{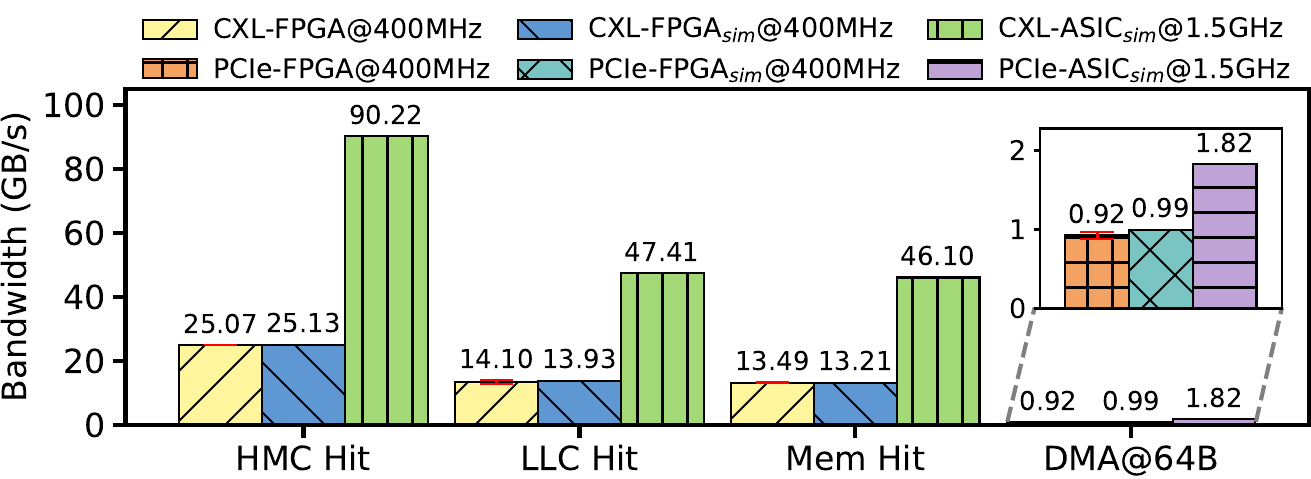}
\caption{Average load bandwidth for cacheline HMC hit, LLC hit, and memory hit via CXL.cache, in comparison to DMA read bandwidth at a \SI{64}{B} message granularity. Error bars represent $\pm 3\sigma$ standard deviation.}
\label{Fig:bandwidth}
\vspace{-10pt}
\end{figure}

\begin{figure}[t]
\centering
\includegraphics[width=0.48\textwidth]{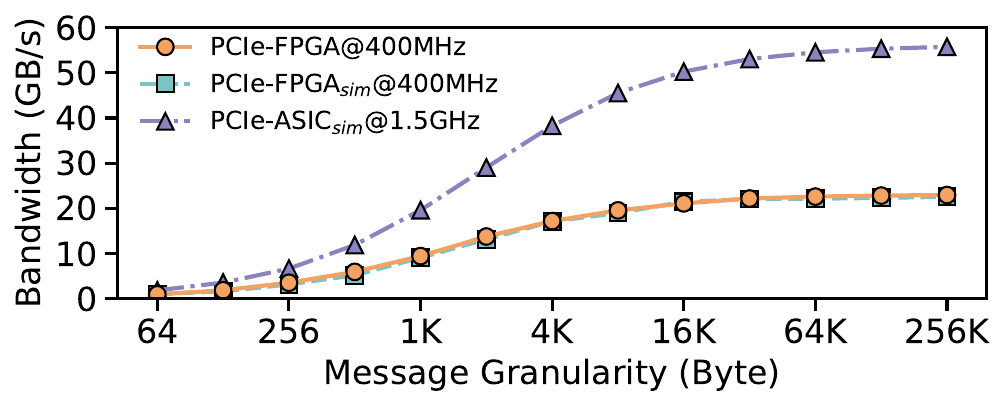}
\caption{Average H2D bandwidth for DMA read with varying message granularity.} 
\label{Fig:dma_bandwidth}
\vspace{-15pt}
\end{figure}

% \vspace{-10pt}
\subsection{SimCXL Bandwidth Calibration}
\label{Sec:SimCXL_Bandwidth_Calibration}
\subsubsection{CXL.cache bandwidth}
To measure the bandwidth of CXL.cache, we configured the LSU to issue a large number of memory requests until a stable bandwidth value was achieved. We observed that when the request count exceeds 512 (totaling \SI{32}{KB}), the bandwidths for HMC hit, LLC hit, and memory hit all converge. We then report the bandwidth results in Fig.~\ref{Fig:bandwidth} at 2,048 requests (totaling \SI{128}{KB}), which are averaged over 1,000 tests. On the 400 MHz CXL-FPGA, the peak bandwidths for HMC hit, LLC hit, and memory hit stabilize at \SI{25.07}{GB/s}, \SI{14.10}{GB/s}, and \SI{13.49}{GB/s}, respectively. Given the FPGA’s theoretical maximum bandwidth of \SI{25.6}{GB/s} at \SI{400}{MHz}, LLC and memory hits reach only 55\% and 52.7\% of the peak bandwidth. This degradation probably stems from pipeline bubbles introduced by cache-coherence checks during requests routed to the host, a behavior also observed in CPU accesses to remote NUMA nodes \cite{demystifying_type3,exploring_cxl_asic}. In contrast, HMC accesses, eliminating the need for host-side coherence checks, achieve 97.7\% of the theoretical bandwidth. One can see that SimCXL accurately reproduces the bandwidth of CXL-FPGA at \SI{400}{MHz}.

\subsubsection{DMA bandwidth}
Fig. \ref{Fig:dma_bandwidth} illustrates the average H2D bandwidth achieved by DMA at different message granularities. On a PCIe-FPGA operating at \SI{400}{MHz}, DMA throughput is very limited for fine-grained transfers. For example, DMA delivers just \SI{0.92}{GB/s} (0.36\% of the theoretical peak) at message size of \SI{64}{B}, whereas CXL.cache achieves \SI{13.25}{GB/s}, 14.4$\times$ that of DMA (see Fig. \ref{Fig:bandwidth}). However, as the message size increases, DMA can pipeline transfers more efficiently, reaching \SI{22.9}{GB/s} at \SI{256}{KB} (86.7\% of the theoretical peak), while CXL.cache achieves only 57.7\% of the DMA bandwidth. These results show that CXL.cache provides a clear throughput advantage for small-message exchanges between host and accelerator, whereas DMA remains the preferred mechanism for bulk transfers. Again, the DMA engine in SimCXL reproduces the FPGA bandwidth curve accurately. Note that after all calibration tests, our simulator achieves a mean absolute percentage error of 3\%.

\begin{figure}[t]
\centering
\includegraphics[width=0.42\textwidth]{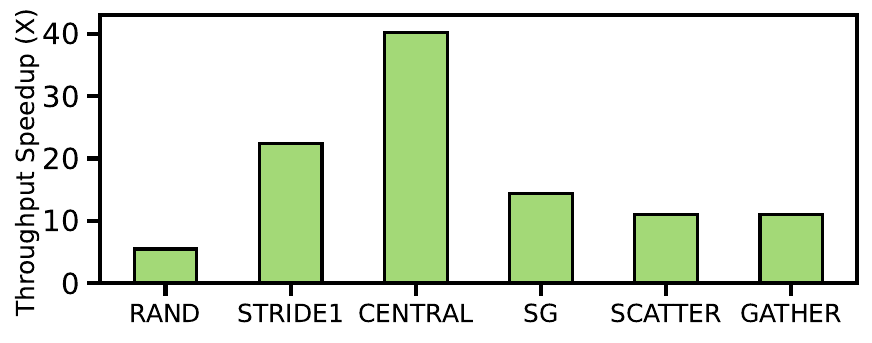}
\caption{Throughput speedup of CXL-based RAO versus PCIe-based RAO across six CircusTent workloads.} 
\label{Fig:rao}
\vspace{-10pt}
\end{figure}

%\vspace{-5pt}
\subsection{Evaluation of Remote Atomic Operations}
We evaluated six RAO access patterns from the CircusTent benchmark \cite{circustent} on both PCIe-NIC and CXL-NIC in SimCXL, with throughput speedups shown in Fig.~\ref{Fig:rao}. The CENTRAL pattern, a many-to-one workload that models a distributed lock service, achieves the highest speedup (40.2$\times$) by caching the hotspot data in CXL-NIC's HMC, avoiding costly PCIe DMA transfers. STRIDE1 follows with a 22.4$\times$ speedup, as its sequential atomic updates on \SI{8}{B} elements leverage \SI{64}{B} cacheline granularity, enabling cache hits for seven subsequent operations after fetching a line. SCATTER and GATHER patterns, which involve randomized updates in the global address space, exhibit moderate speedups due to lower cache hit rates. The SG workload, which combines SCATTER and GATHER, follows a similar trend. Even the fully random pattern RAND, with a near-zero cache hit rate, delivers a 5.5$\times$ throughput gain due to the lower access latency of CXL.cache compared to PCIe-based DMA for fine-grained accesses. These results demonstrate that CXL-NIC significantly enhances the distributed RAO throughput across various access patterns by leveraging low-latency memory access and on-device caching of hot data.

\begin{figure}[t]
\centering
\includegraphics[width=0.5\textwidth]{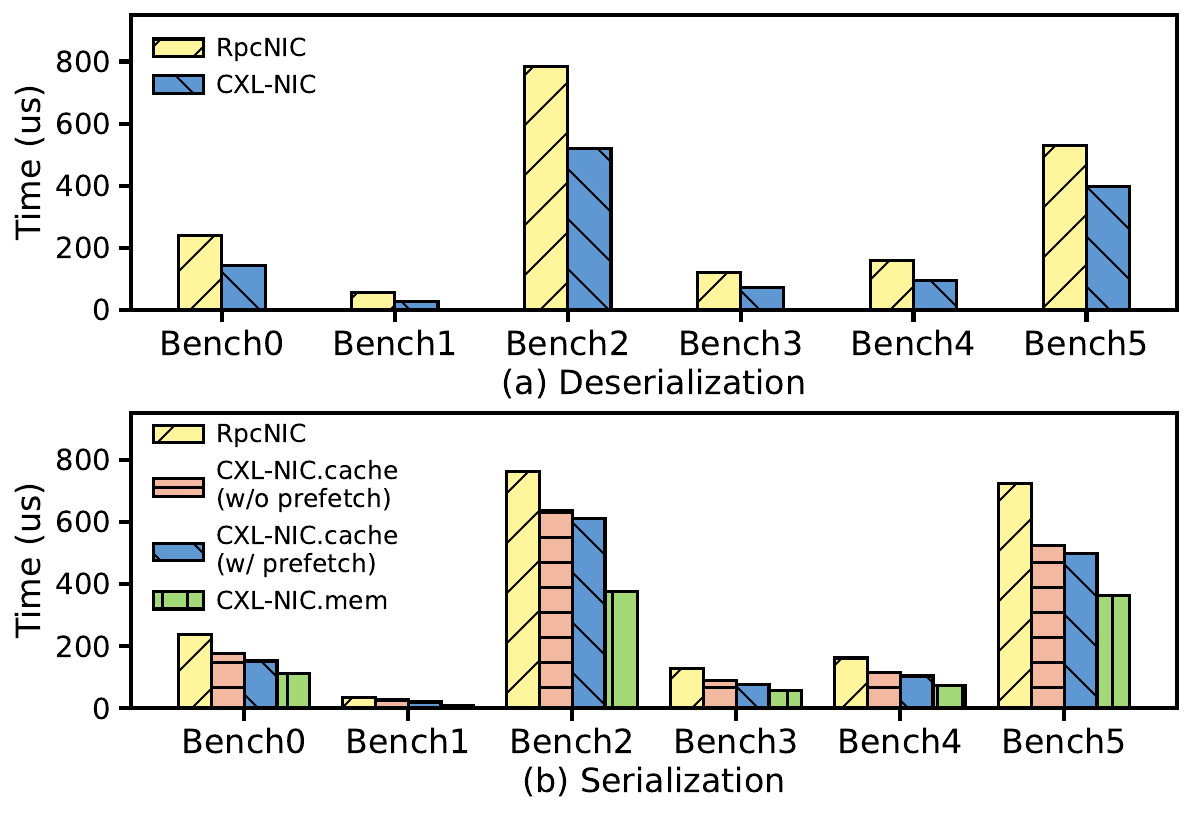}
\caption{De/serialization time of CXL-based RPC versus PCIe-based RPC across six benches from HyperProtoBench.} 
\label{Fig:rpc}
\vspace{-10pt}
\end{figure}

\subsection{Evaluation of Remote Procedure Calls}
We evaluated de/serialization time for CXL-based RPC (CXL-NIC.*) versus PCIe-based RPC (RpcNIC \cite{rpcnic}) using HyperProtoBench \cite{protoAcc}. Fig.~\ref{Fig:rpc}(a) shows deserialization results, with CXL-NIC achieving speedups of 1.33$\times$ (Bench5) to 2.05$\times$ (Bench1) in six benchmarks. This acceleration comes primarily from the CXL-NIC's low-latency fine-grained writes. Although Bench5's large string fields favor DMA, CXL.cache still outperforms RpcNIC by eliminating redundant copies and software overhead. Bench1, with small-field messages, achieves the highest speedup. Moreover, CXL-NIC uses NC-P instructions to push fields directly into the LLC, improving subsequent CPU access performance.

Fig.~\ref{Fig:rpc}(b) presents serialization results, where all three CXL-NIC solutions outperform RpcNIC. Using CXL.mem, the CPU constructs messages directly in device memory, achieving speedups of 2.0$\times$ (Bench5) to 4.06$\times$ (Bench1), as devices access local memory for deserialization, eliminating costly PCIe transfer overhead. Our evaluation results using a Samsung memory expander reveal a 8\% higher overhead at most for message construction through CXL.mem versus construction in host memory, which is an acceptable cost given CXL-NIC's performance gain. With CXL.cache, the CXL-NIC retrieves individual fields directly from host memory, achieving speedups of 1.34$\times$ (Bench2) to 1.65$\times$ (Bench1) when the RPC prefetcher is enabled. 
Compared to the CXL-NIC design without prefetch, our RPC prefetcher improves the serialization performance by 12\% on average across the six benches.
%Moreover, when comparing serialization performance with the prefetcher enabled and disabled, the prefetcher achieves an average improvement of 12\%. 
Note that the minimum gain is 3.6\% for Bench2, since it contains a large number of deeply nested messages limiting prefetch effectiveness. Nonetheless, the CXL-NIC without prefetch still benefits from the low latency of CXL.cache in comparison to RpcNIC. In summary, CXL-based RPCs leverage fine-grained low-latency access and unified memory view to significantly streamline operations, accelerating (de)serialization by an average of 1.86$\times$ when compared to the PCIe-based solution.

\begin{table*}[tb]
\centering
\renewcommand{\arraystretch}{1} 
\caption{Comparison between SimCXL and prior CXL system simulators/emulators.}
\label{tab:SimCXL_vs_others}
\resizebox{1\textwidth}{!}{
\begin{tabular}{|l|c|c|c|c|c|c|c|c|c|}
\hline
\textbf{Simulator/Emulator}  & \makecell{\textbf{Cohet}\\\textbf{Support}}  & \makecell{\textbf{CXL.cache}\\\textbf{Support}}  & \makecell{\textbf{CXL.mem\&io}\\\textbf{Support}} & \makecell{\textbf{CXL XPU}\\\textbf{Models}} & \textbf{Full System} & \makecell{\textbf{Hardware}\\\textbf{Calibration}} & \textbf{Configurability} & \textbf{Sim. Error} & \textbf{Sim. Speed} \\ \hline
{CXLMemSim \cite{CXLMemSim_github}}&{No}&{No}&{No}&{No}&{No}&{No}&{Medium}&{High}&{Medium}\\ \hline
{CXL-DMSim \cite{cxl_dmsim}}&{No}&{No}&{Yes}&{No}&{Yes}&{Yes}&{High}&{Low}&{Low}\\ \hline
{Mess+gem5 \cite{mess}}&{No}&{No}&{No}&{No}&{No}&{No}&{High}&{Medium}&{Low}\\ \hline
{QEMU \cite{QEMU}}&{No}&{No}&{Yes}&{No}&{Yes}&{No}&{High}&{High}&{High}\\ \hline
{Remote NUMA \cite{tarot_row_hammer,dmrpc}}&{No}&{No}&{No}&{No}&{No}&{N/A}&{Low}&{High}&{High}\\ \hline
{SimCXL} &{Yes}&{Yes}&{Yes}&{Yes}&{Yes}&{Yes}&{High}&{Low}&{Low}\\ \hline
\end{tabular}
}
\vspace{-10pt}
\end{table*}

%============================================================================================
\section{Related Work}
\label{Sec:Related_work}
%============================================================================================
\textbf{CXL simulation and emulation methods.} Previous studies often used remote NUMA node (homogeneous in essence) for the emulation of CXL type-1/2 devices \cite{huang_hal_2024,tarot_row_hammer,schuh_cc-nic_2024} or type-3 devices \cite{wahlgren2022evaluating,hpc_cxl_memory,dmrpc,maruf2023tpp}.
Although this approach has greatly facilitated early investigations, the performance characteristics of a remote NUMA node differ substantially from those of a genuine CXL device \cite{demystifying_type2,cxl_memory_character}, which can in some cases lead to misleading conclusions \cite{demystifying_type3}. Existing software simulators focus solely on modeling memory expansion, pooling, and sharing for type-3 devices \cite{QEMU,mess,CXLMemSim_github,cxl_dmsim}, %without providing full-system support for CXL.cache. 
without any touch on heterogeneous systems with typ-1/2 devices.
Table \ref{tab:SimCXL_vs_others} compares SimCXL with other existing CXL system simulation and emulation tools across several dimensions. Only SimCXL models CXL XPU devices and all three CXL sub-protocols, and supports full-system simulation of Cohet computing framework. After meticulous hardware calibration, SimCXL achieves a low simulation error of 3\%.

\textbf{CXL memory disaggregation and pooling.} A wealth of prior work has evaluated real CXL type-3 devices, providing comprehensive performance characterizations and practical insights \cite{demystifying_type3,exploring_cxl_asic,cxl_memory_character,samsung_cxl_ssd}. Many studies have also explored CXL-based tiered memory management at the OS level. \cite{maruf2023tpp,m5,tiered_memory,nomad,neomem}. In addition, some studies exploit CXL memory to expand system memory capacity and bandwidth to accelerate memory-intensive applications \cite{ahnexamination,recomm,hpc_cxl_memory,gpu_graph_process,lia_cxl_llm}. 

\textbf{CXL-based heterogeneous computing.} Recently, a study provides a detailed performance characterization of the CXL type-2 device (FPGA) and offloads two OS kernel memory optimization functions \cite{demystifying_type2}. Some studies leverage the cache coherence and unified memory view provided by CXL.cache to optimize specific application scenarios \cite{huang_hal_2024,tarot_row_hammer,xu_efficient_tensor}. Other works explore host-accelerator synchronization mechanisms based on cache coherence \cite{schuh_cc-nic_2024,mozart,rambda,rag_cxl}. 
%However, these works are limited to applying individual CXL.cache features to specific application scenarios, and their evaluations rely predominantly on  remote NUMA emulation. 
However, these works are limited to: 1) hardware customization exploiting selected CXL.cache features with direct manipulations on physical memory addresses (FPGA-like development),
2) predominately relying on remote NUMA emulation for evaluation.
In contrast, our work builds a Cohet computing framework with a global perspective on the software stack and hardware architecture, greatly simplifying programming model and memory management. We also introduce SimCXL, a research tool that fills a critical gap in CXL-enable heterogeneous systems simulation infrastructure. 
%============================================================================================
\vspace{-10pt}
\section{Discussion and Future Work}
\label{Sec:Discussion_and_Future_Work}
%============================================================================================
Despite its promise, CXL-driven coherent heterogeneous computing still faces many challenges. The overhead of ATS-based address translation remains unexplored due to the lack of ATS support on current CXL FPGA platforms~\cite{no_support_ats}; prior CCIX studies indicate substantial ATC miss penalties~\cite{evaluation_ccix}. Furthermore, OS support for CXL devices remains immature~\cite{cxl_linux}, and mainstream programming frameworks like OpenCL have not yet integrated CXL workflows.

Our future work will focus on adding new features to SimCXL, currently limited to CXL 1.1. With an industry target toward CXL 3.x and its fabric-based multi-node interconnects~\cite{CXL_spec}, we are designing CXL switch models to extend SimCXL and exploring integration with SimBricks~\cite{simbricks} for large-scale fabric-based system simulation. As the coherence domain scales up (i.e., with more child nodes in a supernode), the performance loss and design complexity of maintaining hardware coherence increase sharply, and the huge coherence traffic can become a system bottleneck. To mitigate coherence-traffic storms, we plan to explore a hierarchical coherence protocol for small-scale supernodes. Each child node interacts with a local agent for coherence transactions; the local agent consults a global agent only if it lacks the requested replica. 

Furthermore, we will continue to explore the broader applications of Cohet. In graph processing, large datasets can be stored in a unified memory pool, and graph algorithms with fine-grained random-access patterns \cite{graph1,graph2} offloaded to CXL accelerators can benefit from the coherent CXL interconnect. Similarly, in-memory key-value store operations (e.g., GET/PUT) offloaded to CXL accelerators will benefit from lower-latency, fine-grained memory accesses. Moreover, the superior attributes of CXL such as memory pooling and sharing, p2p direct memory access, and TEE security offer compelling opportunities to accelerate workloads such as large language model inference and distributed services.

%============================================================================================
\section{Conclusion}
\label{Sec:Conclusion}
%============================================================================================
In this work, we introduce Cohet, a CXL-driven coherent heterogeneous computing framework that overcomes the inefficiency of conventional PCIe-based heterogeneous systems in fine-grained host-accelerator interactions and simplifies the heterogeneous programming model. 
We also present SimCXL, a full-system cycle-level simulator that provides a flexible research platform for Cohet computing systems, and it will be open-sourced soon. 
We have rigorously calibrated SimCXL against a real CXL-supported testbed to ensure high-fidelity simulations.
%laying the groundwork for the open-source community to exploit the full potential of coherent heterogeneous computing. 
Finally, we have demonstrated two Cohet-driven killer apps namely RAO and RPC on SimCXL, demonstrating significant performance gains and streamlined design flows compared to PCIe-based solutions.

\section*{Acknowledgement}
We would like to thank the anonymous reviewers for their insightful and constructive feedback. This work was supported in part by the National Key Research and Development Program of China under Grant 2022YFB4500304, and in part by the National Natural Science Foundation of China under Grants 62332021, 62304257 and 62472058.

%%%%%%% -- PAPER CONTENT ENDS -- %%%%%%%%

%%%%%%%%% -- BIB STYLE AND FILE -- %%%%%%%%
\balance
\bibliographystyle{IEEEtran}
\bibliography{refs}

%\bibliographystyle{unsrt}
%%%%%%%%% -- Appendix -- %%%%%%%%%
% \include{ae-20201122}

\end{document}

%% file: hpca-template.tex
%%%%%%%%%%%%%%%%%%%%%%%%%%%%%%%%%%%%%
%%%%%%%%%% -- DO NOT MODIFY -- %%%%%%%%%%
%%%%%%%%%%%%%%%%%%%%%%%%%%%%%%%%%%%%%

\author{
  \ifdefined\hpcacameraready
    \IEEEauthorblockN{\hpcaauthors{}}
      \IEEEauthorblockA{
        \hpcaaffiliation{} \\
        \hpcaemail{}
      }
  \else
    \IEEEauthorblockN{\normalsize{HPCA \hpcayear{} Submission
      \textbf{\#\hpcasubmissionnumber{}}} \\
      \IEEEauthorblockA{
        Confidential Draft \\
        Do NOT Distribute!!
      }
    }
  \fi 
}

% Heading and footer for title page
\fancypagestyle{camerareadyfirstpage}{%
  \fancyhead{}
  \renewcommand{\headrulewidth}{0pt}
  \fancyhead[C]{
    \ifdefined\aeopen
    \parbox[][12mm][t]{13.5cm}{\hpcayear{} IEEE International Symposium on High-Performance Computer Architecture (HPCA)}    
    \else
      \ifdefined\aereviewed
      \parbox[][12mm][t]{13.5cm}{\hpcayear{} IEEE International Symposium on High-Performance Computer Architecture (HPCA)}
      \else
      \ifdefined\aereproduced
      \parbox[][12mm][t]{13.5cm}{\hpcayear{} IEEE International Symposium on High-Performance Computer Architecture (HPCA)}
      \else
      \parbox[][0mm][t]{13.5cm}{\hpcayear{} IEEE International Symposium on High-Performance Computer Architecture (HPCA)}
    \fi 
    \fi 
    \fi 
    \ifdefined\aeopen 
      \includegraphics[width=12mm,height=12mm]{ae-badges/open-research-objects.pdf}
    \fi 
    \ifdefined\aereviewed
      \includegraphics[width=12mm,height=12mm]{ae-badges/research-objects-reviewed.pdf}
    \fi 
    \ifdefined\aereproduced
      \includegraphics[width=12mm,height=12mm]{ae-badges/results-reproduced.pdf}
    \fi
  }
  %\fancyfoot[L]{\hpcapubid{} \copyright \hpcayear{} IEEE}
  \fancyfoot[C]{}
}
% Heading and footer for remaining pages
\fancyhead{}
\renewcommand{\headrulewidth}{0pt}
%\fancyhead[C]{\hpcayear{} IEEE International Symposium on
% High-Performance Computer Architecture (HPCA)}

%Enables the camera ready header and footer
\ifdefined\hpcacameraready 
  \thispagestyle{camerareadyfirstpage}
  \pagestyle{empty}
\else
  \thispagestyle{plain}
  \pagestyle{plain}
\fi

\newcommand{\hpcaheight}{0mm}
\ifdefined\eaopen
\renewcommand{\hpcaheight}{12mm}
\fi